\documentclass[12pt]{article}
\usepackage{amssymb,amscd,array}
\catcode `\@=11
\@addtoreset{equation}{subsection}

\def\harr#1#2{\smash{\mathop{\hbox to .5in{\rightarrowfill}}
\limits^{\scriptstyle#1}_{\scriptstyle#2}}}

\def\harrl#1#2{\smash{\mathop{\hbox to .5in{\leftarrowfill}}
\limits^{\scriptstyle#1}_{\scriptstyle#2}}}

\def\qed{\blacksquare}
\newcommand{\be}{\begin{equation}}
\newcommand{\ee}{\end{equation}}
\newcommand{\R}{\mathbb{R}}
\newcommand{\N}{\mathbb{N}}
\newcommand{\C}{\mathbb{C}}
\newtheorem{thm}{Theorem}[section]
\newtheorem{rem}[thm]{Remark}
\newtheorem{lemma}[thm]{Lemma}
\newtheorem{cor}[thm]{Corollary}
\newtheorem{prop}[thm]{Proposition}
\begin{document}
\begin{titlepage}
\begin{center}
{\bf \Large{On the Uniqueness of the Non-Abelian Gauge Theories ~\\
in Epstein-Glaser Approach to Renormalisation Theory\\}}
\end{center}
\vskip 1.0truecm
\centerline{D. R. Grigore
\footnote{e-mail: grigore@theor1.ifa.ro, grigore@roifa.ifa.ro}}
\vskip5mm
\centerline{Dept. of Theor. Phys., Inst. Atomic Phys.}
\centerline{Bucharest-M\u agurele, P. O. Box MG 6, ROM\^ANIA}
\vskip 2cm
\bigskip \nopagebreak
\begin{abstract}
\noindent
We generalise a result of Aste and Scharf saying that, under some reasonable
assumptions, consistent with renormalisation theory, the non-Abelian gauge
theories describes the only possibility of coupling the gluons. The proof is
done using Epstein-Glaser approach to renormalisation theory.

\end{abstract}
\end{titlepage}

\section{Introduction}

Renormalisation theory has a long history and among the many achievements one
can count quantum electrodynamics (with the extremely good experimental
verification) and, more recently, the non-Abelian gauge theories which are used
to describe electro-weak theory and quantum chromodynamics. The major point is
that these theories are renormalisable i.e. one can get rid of the so-called
ultra-violet divergences in a consistent way and obtain a well defined series
for the $S$-matrix. Using the traditional approach based on Feynman diagrams,
this assertion has been proved in second order of the perturbation theory by 't
Hooft and Veltman and has been established in all orders of the perturbation
theory in the extremely sophisticated approach of Becchi, Rouet and Stora.

However, it was proved by Epstein and Glaser \cite{EG1}, \cite{Gl} that the
most natural and straightforward way of dealing with perturbation theory and
constructing a $S$-matrix fulfilling Bo\-go\-liu\-bov axioms, is based on a
direct exploitation of the causality axiom, and this can be done using a
technical device, called the distribution splitting. This analysis was
performed for a scalar field and extended for quantum electrodynamics in an
external field in \cite{DM}. The full analysis of the interacting QED was done
by G. Scharf and collaborators and is presented in a pedagogical manner in
\cite{Sc1}. In recent years, the analysis was extended to the case of pure
non-Abelian gauge symmetries \cite{DHKS1}, \cite{DHKS2}, \cite{DHS2},
\cite{DHS3}, \cite{Hu1}-\cite{Hu4}, \cite {Kr1}, \cite{Kr3} and of the
electro-weak theory \cite{DS}, \cite{ASD3}. Many other achievements of this
approach are illustrated by the bibliography. 

Recently Aste and Scharf \cite{AS} proved that the only possibility of
non-trivial coupling between $r > 1$ zero-mass vector fields is through the
usual Yang-Mills recipe.  In other words, the existence of a compact
semi-simple Lie group (to which the gauge principle is applied to obtain a
local gauge theory) should not be assumed from the beginning: it simply follows
from the consistency requirements of the theory.  The main hypothesis leading
to this result was a certain quantum form of the gauge principle invariance
imposed to the $S$-matrix. 

In this paper we will generalise this result and simplify somewhat they proof.
Let us explain in what sense our approach is more general. The main obstacle in
constructing the perturbation series for a zero-mass field is the fact that, as
it happens for the electromagnetic field, one is forced to use non-physical
degrees of freedom for the description of the free fields \cite{WG},
\cite{SW1}, \cite{Ka} in a Fock space formalism; one introduces in this Fock
space a non-degenerate sesquilinear form and the true physical Hilbert space is
obtained by a certain process of factorisation. More sophisticated, one can
extend the Fock space to an auxiliary Hilbert space ${\cal H}_{gh}$ including
some fictious fields, called ghosts, and construct an supercharge (i.e. an
operator $Q$ verifying $Q^{2} = 0$) such that the physical Hilbert space is 
$
{\cal H}_{phys} \equiv {\it Ker}(Q)/{\it Im}(Q)
$ 
(see for instance
\cite{We} and references quoted there).  We will present a careful analysis
(which seems to be missing from the literature) of the Fock space
representation of the electromagnetic field.  The problem that one faces in the
attempt to construct the perturbative $S$ matrix {\it \'a la} Epstein-Glaser is
that one can define Wick monomials (from which the $S$-matrix is built) only on
the auxiliary Fock space 
${\cal H}_{gh}$; 
so one must impose, beside the usual Bogoliubov axioms, the supplementary
condition that the $S$ matrix factorizes to 
${\cal H}_{phys}$.  
We will prove that, the combination of these conditions leads uniquely to the
pure Yang-Mills interaction. No other hypothesis are necessary to prove this
statement. It is necessary to study rigorously the $S$ matrix only up to
order $2$ in the perturbative sense. The gauge invariance condition is a
consequence of our analysis and it is not necessary to impose it as an
independent axiom. We also hope to present a simpler approach to the question
of unitarity of the $S$-matrix.

The same analysis will be then applied to a theory of quarks and gluons. The
result is that the Dirac Fermions should form a multiplet of the semi-simple
compact group which appears naturally from the analysis of the pure Yang-Mills
case. 

We mention here that results of these type have been obtained quite a long time
ago in \cite{Le} and \cite{CLT} from some arguments concerning the high energy
behaviour of the $S$ matrix elements in the tree approximation. However, a
rigorous basis of this analysis seems to be lacking.

In Section 2 we will present the general scheme of construction of a
perturbation theory in the causal approach of Epstein and Glaser giving all the
relevant details about the Fock space construction of the theory. In Section 3
we will study in detail the quantization of the electromagnetic field, because
some subtle points are simply overlooked in the literature or presented too
summarily and not completely correct. We hope that we will be able to argue
convincingly that a deep an clear understanding of these points is essential
for the understanding of quantum gauge theories. In Section 4 we present our
main result concerning the unicity of the Yang-Mills interaction and we
generalise the analysis to the case when matter fields (quarks) are present. In
the last Section we indicate some further possible developments.

Regarding the level of rigor, we make the following comments: (a) we will work
with the formal notations for distributions for reasons of simplicity; (b) when
working with Hilbert spaces of 
$L^{2}(X,d\alpha)$ 
one has to consider not functions but classes of function which are identically
almost everywhere (a.e.); (c) domain problems for unbounded operators can be
fixed in standard ways \cite{WG}, \cite{RR}.
\newpage

\section{Perturbation Theory in Fock Spaces}

\subsection{Second Quantization\label{sq}}

Here we give the main concepts and formul\ae~ connected to the method of second
quantization. We follow essentially \cite{Va} ch. VII; more details can be found
in \cite{Co} and \cite{Be}. 

The idea of the method of second quantization is to provide a canonical
framework for a multi-particle system in case one has a Hilbert space 
describing an ``elementary" particle. (One usually takes the one-particle
Hilbert space 
${\sf H}$
to be some projective unitary irreducible representation of the Poincar\'e 
group, but this is not important for this subsection). Let 
${\sf H}$ 
be a (complex) Hilbert space; the scalar product on {\sf H} is denoted by 
$<\cdot,\cdot>$.
One first considers the tensor algebra
\be
T({\sf H}) \equiv \oplus_{n=0}^{\infty} {\sf H}^{\otimes n},
\ee
where, by definition, the term corresponding to 
$n = 0$
is the division field $\C$. 
The generic element of 
$T({\sf H})$
is of the type
$
(c,\Phi^{(1)},\cdots,\Phi^{(n)},\cdots), \Phi^{(n)} \in {\sf H}^{\otimes n};
$
the element
$\Phi_{0} \equiv (1,0,\dots)$
is called {\it the vacuum}.
Let us consider now the symmetrisation (resp. antisymmetrisation) operators
${\cal S}^{\pm}$
defined by
\be
{\cal S}^{\pm} \equiv \oplus_{n=0}^{\infty} {\cal S}_{n}^{\pm}
\ee
where 
${\cal S}_{0}^{\pm} = 1$
and
${\cal S}_{n}^{\pm}, \quad n \geq 1$
are defined on decomposable elements in the usual way
\be
{\cal S}_{n}^{+} \phi_{1} \otimes \cdots \otimes \phi_{n} \equiv
{1\over n!} \sum_{P \in {\cal P}_{n}} 
\phi_{P(1)} \otimes \cdots \otimes \phi_{P(n)}
\ee
and
\be
{\cal S}_{n}^{-} \phi_{1} \otimes \cdots \otimes \phi_{n} \equiv
{1\over n!} \sum_{P \in {\cal P}_{n}} (-1)^{|P|}
\phi_{P(1)} \otimes \cdots \otimes \phi_{P(n)};
\ee
here 
${\cal P}_{n}$
is the group of permutation of the numbers 
$1,2,\dots,n$
and
$|P|$
is the sign of the permutation $P$. One extends the operators 
${\cal S}_{n}^{\pm}$
arbitrary elements of $T$ by linearity and continuity; it is convenient to
denote the elements in defined by these relations by
$\phi_{1} \vee \cdots \vee \phi_{n}$
and respectively by
$\phi_{1} \wedge \cdots \wedge \phi_{n}$.

We now define the 
{\it Bosonic} (resp. {\it Fermionic}) {\it Fock space} according to:
\be
{\cal F}^{\pm}({\sf H}) \equiv {\cal S}^{\pm} T({\sf H}); 
\ee
obviously we have:
\be
{\cal F}^{\pm}({\sf H}) = \oplus_{n=0}^{\infty} {\cal H}^{\pm}_{n}
\ee
where
\be
{\cal H}^{\pm}_{0} \equiv \C, \quad 
{\cal H}^{\pm}_{n} \equiv {\cal S}_{n}^{\pm} {\sf H}^{\otimes n}
\quad (n \geq 1)
\ee
are the so-called $n^{\rm th}$-{\it particle subspaces}.

The operations $\vee$ (resp. $\wedge$) make
${\cal F}^{\pm}({\sf H})$
into associative algebras. One defines in the Bosonic (resp. Fermionic) Fock
space the {\it creation} and {\it annihilation} operators as follow: let
$\phi \in {\sf H}$ be arbitrary.
In the Bosonic case they are defined on elements from 
$\psi \in {\cal H}^{+}_{n}$ by
\be
A(\phi)^{\dagger} \psi \equiv \sqrt{n+1} \phi \vee \psi
\ee
and respectively
\be
A(\phi) \psi \equiv {1 \over \sqrt{n}} i_{\phi} \psi
\ee
where
$i_{\phi}$
is the unique derivation of the algebra
${\cal F}^{+}({\sf H})$
verifying
\be
i_{\phi} 1 = 0; \quad i_{\phi} \psi = <\phi,\psi> {\bf 1}.
\ee

\begin{rem}
We note that the general idea is to associate to every element of the
one-particle space 
$\phi \in {\sf H}$ 
a couple of operators 
$A^{\sharp}(\phi)$
acting in the Fock space
${\cal F}^{+}({\sf H})$.
\label{creation+annihilation}
\end{rem}

As usual, we have the canonical commutation relations (CCR):
\be
\left[ A(\phi), A(\psi)\right] = 0,\quad
\left[ A(\phi)^{\dagger}, A(\psi)^{\dagger}\right] = 0, \quad
\left[ A(\phi), A(\psi)^{\dagger}\right] = <\phi,\psi> {\bf 1}.
\label{CCR}
\ee

The operators
$A(\psi), \quad A(\psi)^{\dagger}$
are unbounded and adjoint one to the other.

In the Fermionic case we define these operators on elements from 
$\psi \in {\cal H}^{-}_{n}$ by
\be
A(\phi)^{\dagger} \psi \equiv \sqrt{n+1} \phi \wedge \psi
\ee
and respectively
\be
A(\phi) \psi \equiv {1 \over \sqrt{n}} i_{\phi} \psi
\ee
where
$i_{\phi}$
is the unique graded derivation of the algebra
${\cal F}^{-}({\sf H})$
verifying
\be
i_{\phi} 1 = 0; \quad i_{\phi} \psi = <\phi,\psi> {\bf 1}.
\ee

Now we have the canonical anticommutation relations (CAR):
\be
\left\{ A(\phi), A(\psi)\right\} = 0,\quad
\left\{ A(\phi)^{\dagger}, A(\psi)^{\dagger}\right\} = 0, \quad
\left\{ A(\phi), A(\psi)^{\dagger}\right\} = <\phi,\psi> {\bf 1}.
\label{CAR}
\ee

The operators
$A(\psi), \quad A(\psi)^{\dagger}$
are bounded and adjoint one to the other.

If $U$ is a unitary (or antiunitary) operator on ${\sf H}$, 
it lifts naturally to an operator 
$\Gamma(U)$
on the tensor algebra 
$T({\sf H})$, 
according to
\be
\Gamma(U) \equiv \oplus_{0}^{\infty} U^{\otimes n}
\ee
or, more explicitly, on decomposable elements
\be
\Gamma(U) \psi_{1} \otimes \cdots \otimes \psi_{n} = 
U \psi_{1} \otimes \cdots \otimes U \psi_{n}.
\ee

The operator
$\Gamma(U)$
leaves invariant the symmetric and resp. the antisymmetric algebras
${\cal F}^{\pm}({\sf H})$
and we have
\be
\Gamma(U_{g}) A(\phi) \Gamma(U_{g^{-1}}) = A(U_{g}\phi).
\label{repsA}
\ee

\newpage

\subsection{Elementary Relativistic Free Particles\label{erp}}

As we have anticipated in the previous subsection, on usually takes
${\sf H}$ to be the Hilbert space of an unitary irreducible representation of
the Poincar\'e group. We give below the relevant formul\ae~ for the scalar
particle of mass $m$ and for the photon. By comparison, one will be able to see
the origin of the difficulties of the renormalisation theory for zero-mass
particles. 

According to \cite{Va}, a scalar particle of mass $m$ can be described in the
Hilbert space
${\sf H} \equiv L^{2}(X^{+}_{m},\C,d\alpha_{m}^{+})$
of Borel complex function 
$\phi$
defined on the upper hyperboloid of mass 
$m \geq 0$
$X^{+}_{m} \equiv \{p \in \R^{4} \vert \quad \Vert p\Vert^{2} = m^{2}\}$
which are square integrable with respect to the Lorentz invariant measure
$d\alpha_{m}^{+} \equiv {d{\bf p} \over 2\omega({\bf p})}$.
Here the conventions are the following:
$\Vert \cdot \Vert$
is the Minkowski norm defined by
$
\Vert p \Vert^{2} \equiv p\cdot p 
$
and
$p\cdot q$
is the Minkowski bilinear form:
\be
p\cdot q \equiv p_{0}q_{0} - {\bf p} \cdot {\bf q}.
\ee

If
${\bf p} \in \R^{3}$
we define
$\tau({\bf p}) \in X_{m}^{+}$
according to
$
\tau({\bf p}) \equiv (\omega({\bf p}),{\bf p}), \quad 
\omega({\bf p}) \equiv \sqrt{{\bf p}^{2} + m^{2}}.
$

The scalar product in 
${\sf H}$
is:
\be
<\phi,\psi> \equiv \int_{X^{+}_{m}} d\alpha_{m}^{+} \overline{\phi(p)} \psi(p).
\ee

The expression for the corresponding unitary irreducible representation of the
Poincar\'e group is:
\be
\left( U_{a,\Lambda} \phi\right)(p) \equiv 
e^{ia\cdot p} \phi(\Lambda^{-1}\cdot p)\quad {\rm for} \quad \Lambda \in 
{\cal L}^{\uparrow}, \qquad 
\left( U_{I_{t}} \phi\right)(p) \equiv \overline{ \phi(I_{s}\cdot p)};
\ee
here 
$I_{s}, \quad I_{t}$
are the elements of the Lorentz group corresponding to the spatial and 
respectively temporal inversion. Also by
$(\Lambda,p) \mapsto \Lambda \cdot p$ 
we denote the usual action of the Lorentz group on $\R^{4}$ and $\C^{4}$.
The couple 
$({\sf H},U)$
is called {\it scalar particle}.

For the photon, such a simple description of the Hilbert space as a space of
functions is no longer available. However, we can obtain a description of this
type if one considers a factorisation procedure \cite{Va}. Let us consider the 
Hilbert space
${\sf H} \equiv L^{2}(X^{+}_{0},\C^{4},d\alpha_{m}^{+})$
with the scalar product
\be
<\phi,\psi> \equiv \int_{X^{+}_{0}} d\alpha_{0}^{+} <\phi(p),\psi(p)>_{\C^{4}}
\ee
where
$
<u,v>_{\C^{4}} \equiv \sum_{i=1}^{4} \overline{u_{i}} v_{i}
$
is the usual scalar product from $\C^{4}$. In this Hilbert space we have the
following (non-unitary) representation of the Poincar\'e group:
\be
\left( U_{a,\Lambda} \phi\right)(p) \equiv 
e^{ia\cdot p} \Lambda \cdot \phi(\Lambda^{-1}\cdot p)\quad {\rm for} \quad 
\Lambda \in {\cal L}^{\uparrow}, \qquad 
\left( U_{I_{t}} \phi\right)(p) \equiv \overline{ \phi(I_{s}\cdot p)}.
\label{reps0}
\ee

Let us define on ${\sf H}$ the operator $g$ by
\be
(g\cdot \phi)(p) \equiv g\cdot \phi(p)
\label{g}
\ee
and following non-degenerate sesquilinear form:
\be
(\phi,\psi) \equiv - <\phi,g\cdot \psi>;
\ee
here 
$g \in {\cal L}^{\uparrow}$
is the Minkowski matrix with diagonal elements
$1,-1,-1,-1$
and the operator $g$ is appearing in (\ref{g}) also called a {\it Krein
operator}. Explicitly:
\be
(\phi,\psi) \equiv \int_{X^{+}_{0}} d\alpha_{0}^{+} (\phi(p),\psi(p)) =
\int_{X^{+}_{0}} d\alpha_{0}^{+} g^{\mu\nu} 
\overline{\phi_{\mu}(p)}\psi_{\nu}(p);
\ee
the indices $\mu,\nu$
take the values 
$0,1,2,3$
and the summation convention over the dummy indices is used.

Then one easily establishes that we have
\be
(U_{a,\Lambda} \phi,U_{a,\Lambda} \psi) = (\phi,\psi), \quad {\rm for} \quad 
\Lambda \in {\cal L}^{\uparrow}, \qquad 
(U_{I_{t}} \phi,U_{I_{t}} \psi) = \overline{(\phi,\psi)}.
\label{reprezentation}
\ee

We have now two elementary results:
\begin{lemma}
Let us consider the following subspace of ${\sf H}$:
\be
{\sf H}'\equiv \{ \phi \in {\sf H} \vert \quad p^{\mu} \phi_{\mu}(p) = 0\}.
\ee

Then the sesquilinear form
$\left. (\cdot,\cdot)\right|_{\sf H}$
is positively defined.
\label{h'}
\end{lemma}

\begin{lemma}
Let us consider the following subspace of ${\sf H}'$:
\be
{\sf H}''\equiv \{ \phi \in {\sf H}' \vert \quad \Vert\phi\Vert = 0\}.
\ee

Then
\be
{\sf H}''\equiv \{ \phi \in {\sf H} \vert \quad  {\rm there~ exists} \quad
\lambda: X^{+}_{0} \rightarrow \C\quad {\rm s.t.} 
\quad \phi(p) = p \lambda(p)\}.
\ee
\label{h''}
\end{lemma}

Then we have the following result:
\begin{prop}
The representation (\ref{reprezentation}) of the Poincar\'e group leaves
invariant the subspaces ${\sf H}'$ and ${\sf H}''$ and so, it induces an
representation in the Hilbert space
\be
{\rm H}_{photon} \equiv \overline{\left( {\sf H}' / {\sf H}''\right)}
\label{photon}
\ee
(here by the overline we understand completion). The factor representation,
denoted also by $U$ is unitary and irreducible. By restriction to the proper
orthochronous Poincar\'e group it is equivalent to the representation 
${\sf H}^{[0,1]} \oplus {\sf H}^{[0,-1]}$.
\end{prop}

By definition, the couple
$({\rm H}_{photon},U)$
is called {\it photon}. 
\newpage
\subsection{Free Fields and Wick products\label{ff}}

Let us apply the second quantization procedure to the scalar particle, i.e. we
consider that, in the general scheme from the first subsection, the
one-particle subspace ${\sf H}$ is the Hilbert space corresponding to the
scalar particle and we consider Bose statistics i.e. the Bosonic Fock space. 
Then one can canonically identify the 
$n^{\rm th}$-particle subspace
$
{\cal F}^{+}({\sf H})
$
with the set of Borel functions
$\Phi^{(n)}: (X^{+}_{m})^{\times n} \rightarrow \C$
which are square integrable with respect to the product measure
$d(\alpha_{m}^{+})^{\times n}$
and verify the symmetry property
\be
\Phi^{(n)}(p_{P(1)},\dots,p_{P(n)}) = \Phi^{(n)}(p_{1},\dots,p_{n}), \quad 
\forall P \in {\cal P}_{n}.
\label{sym-scalar}
\ee

On the dense domain of test functions one can define the annihilation
operators: 
\be
\left( A(k) \Phi\right)^{(n)} (p_{1},\dots,p_{n}) \equiv \sqrt{n+1}
\Phi^{(n+1)}(k,p_{1},\dots,p_{n}).
\label{annihilation}
\ee

Working with the formalism of rigged Hilbert spaces one can also make sense of
the creation operators:
\be
\left( A(k)^{\dagger} \Phi\right)^{(n)} (p_{1},\dots,p_{n}) \equiv 
2 \omega({\bf k}) {1\over \sqrt{n}} \sum_{i=1}^{n} \delta(k-p_{i})
\Phi^{(n-1)}(p_{1},\dots,\hat{p_{i}},\dots,p_{n})
\label{creation}
\ee
where the Bourbaki conventions
$\sum_{\emptyset} \equiv 0, \quad \prod_{\emptyset} \equiv 1$
are used.

These operators verify the canonical commutation relations (see (\ref{CCR})):
\be
\left[ A(k), A(k')\right] = 0,\quad
\left[ A(k)^{\dagger}, A(k')^{\dagger}\right] = 0, \quad
\left[ A(k), A(k')^{\dagger}\right] = 2 \omega({\bf k}) \delta{(k-k')} {\bf 1}
\label{CCR-em}
\ee
and can be used to express the creation and annihilation operators
$A(\phi), \quad A(\phi)^{\dagger}$
defined in the preceding section; namely we have:
\be
A(\phi) = \int_{X^{+}_{m}} d\alpha^{+}_{m}(k) \overline{\phi(k)} A(k), \quad
A(\phi)^{\dagger} = \int_{X^{+}_{m}} d\alpha^{+}_{m}(k) \phi(k) A(k)^{\dagger}.
\ee

Now we define the scalar free field. Let 
$f \in {\cal S}(\R^{4})$
be a test function. We define the Fourier transform with the convention:
\be
\tilde{f}(p) \equiv {1\over (2\pi)^{2}} \int_{\R^{4}} e^{-ip\cdot x} f(x).
\ee
We also define the functions
\be
f_{\pm}(k) \equiv \left. \tilde{f}(\pm k)\right|_{X^{+}_{m}}.
\ee

One can give now the expression of the {\it free scalar field of mass} $m$ on
the dense domain of the test functions according to:
\begin{eqnarray}
\left( \varphi(f)\Phi\right)^{(n)}(p_{1},\cdots,p_{n}) = \sqrt{2\pi} 
[ \sqrt{n+1} \int_{X^{+}_{m}} d\alpha^{+}_{m}(k)
f_{+}(k) \Phi^{(n+1)}(k,,p_{1},\cdots,p_{n}) + 
\nonumber \\
{1\over \sqrt{n}} \sum_{i=1}^{n} f_{-}(p_{i})
\Phi^{(n-1)}(p_{1},\cdots,\hat{p_{i}},\cdots,p_{n}) ].
\label{free-scalar}
\end{eqnarray}

Then one can extend it to a selfadjoint operator on the Fock space. 
One can define the scalar field in a fixed point as follows. First one defines
the {\it negative frequency part} by:
\be
\left( \varphi_{-}(x)\Phi\right)^{(n)}(p_{1},\cdots,p_{n}) \equiv
(2\pi)^{-3/2} \sqrt{n+1} \int_{X^{+}_{m}} d\alpha^{+}_{m}(k)
e^{-ix\cdot k} \Phi^{(n+1)}(k,,p_{1},\cdots,p_{n}) 
\ee
as a legitimate operator in the Fock space. Working with rigged Hilbert spaces
one can define the {\it positive frequency part}:
\be
\left( \varphi_{+}(x)\Phi\right)^{(n)}(p_{1},\cdots,p_{n}) \equiv
(2\pi)^{-3/2} {1\over \sqrt{n}} \sum_{i=1}^{n} e^{ix\cdot p_{i}}
\Phi^{(n-1)}(p_{1},\cdots,\hat{p_{i}},\cdots,p_{n}).
\ee

Then the expression
\be
\varphi(x) \equiv \varphi_{+}(x) + \varphi_{-}(x)
\ee
is the called {\it real scalar field in the point} $x$. One can justify this 
definition by making sense of the formula:
\begin{eqnarray}
\varphi(f) = \int_{\R^{4}} dx f(x) \varphi(x) =
\sqrt{2\pi} \int_{X^{+}_{0}} d\alpha^{+}_{m}(k)
\left[f_{+}(k) A(k) + f_{-}(k) A^{\dagger}(k\right] =
\nonumber \\
\sqrt{2\pi} \left[ A(\overline{f_{+}}) + A^{\dagger}(f_{-})\right].
\label{free-scalar1}
\end{eqnarray}

The
attribute {\it real} is due to the equation:
\be
\varphi(x)^{*} = \varphi(x),
\ee
the
attribute {\it free} to the equation:
\be
\square \varphi(x) = 0
\ee
and the attribute {\it scalar} to the transformation properties with respect to
the Poincar\'e group: if we define the natural extension of the representation
(\ref{reprezentation}) to the Fock space:
\be
{\cal U}_{g} \equiv \Gamma(U_{g}), \quad \forall g \in {\cal P}
\ee
then we have:
\begin{eqnarray}
{\cal U}_{a,\Lambda} \varphi(x) {\cal U}_{a,\Lambda}^{-1} = 
\varphi(\Lambda\cdot x +a),  \quad \forall \lambda \in {\cal L}^{\uparrow},
\quad
{\cal U}_{I_{t}} \varphi(x) {\cal U}_{I_{t}} = \varphi(I_{s}\cdot x)
\end{eqnarray}

Moreover, we have the important {\it causality} property:
\be
\left[ \varphi(x), \varphi(y)\right] = 0, 
\quad {\rm for} \quad (x-y)^{2} \leq 0.
\label{causal}
\ee

An important generalisation of the notion of free field is given by the concept
of Wick monomials. According to the rigourous treatment of \cite{WG} one can
make sense of the following expressions:
\be
W_{rs}(x) \equiv \varphi_{+}(x)^{r} \varphi_{-}(x)^{s}, \quad r,s \in \N.
\ee

Indeed, if one formally integrates with a test function 
$f \in {\cal S}(\R^{4})$
the expression defined above, one gets the following expression:
\be
W_{rs}(f) \equiv \int_{\R^{4}} f(x) W_{rs}(x)
\ee
for which the following explicit formul\ae~ are available. For $n < r$:
\be
\left( W_{rs}(f) \Phi\right)^{(n)} = 0 
\ee
and for $n \geq r$:
\begin{eqnarray}
\left( W_{rs}(f) \Phi\right)^{(n)} = (2\pi)^{2-{3\over 2}(r+s)} 
{\sqrt{(n-r+s)!n!} \over (n-r)!} \times
\nonumber \\
{\cal S}_{n} \int_{(X^{+}_{m})^{\times s}} 
\prod_{j=1}^{s} d\alpha^{+}_{m}(k_{j})
\tilde{f}\left( \sum_{j=1}^{s} \tau({\bf k}_{j}) - 
\sum_{i=1}^{r} \tau({\bf p}_{i})\right) 
\Phi^{(n-r+s)}(k_{1},\cdots,k_{s},p_{r+1},\cdots,p_{n}); 
\end{eqnarray}
here the operator
${\cal S}_{n}$
symmetrizes in the variables
$p_{1},\cdots,p_{n}.$

The central result making the expressions above legitimate operators in the
Fock space is the following lemma \cite{SW1}.

\begin{lemma}
In the conditions above the following function
\be
F(p_{1},\cdots,p_{n}) \equiv
\int_{(X^{+}_{m})^{\times s}} \prod_{j=1}^{s} d\alpha^{+}_{m}(k_{j})
\tilde{f}\left( \sum_{j=1}^{s} \tau({\bf k}_{j}) - 
\sum_{i=1}^{r} \tau({\bf p}_{i})\right) 
\Phi^{(n-r+s)}(k_{1},\cdots,k_{s},p_{r+1},\cdots,p_{n})
\label{wick}
\ee
is a test function.
\label{Wick-prod}
\end{lemma}

Now one defines {\it Wick products} to be the expressions
\be
:\varphi^{l}:(f) \equiv \sum_{r+s=l} W_{rs}(f) = \int_{\R^{4}} f(x)
:\varphi(x)^{l}:
\ee
with the usual definition for
$:\varphi(x)^{l}:$
namely one puts all the creation operators at the left of the annihilation
operators. In a similar way one can define Wick product with some derivative on
the various factors from 
$:\varphi(x)^{l}:$.

Now we remind Wick theorem. Suppose that 
$A_{1}(x),\dots,A_{n}(x),B_{1}(x),\dots,B_{m}(x)$
are the free scalar field or derivatives of it. Then we have the following
formula: 
\begin{eqnarray}
:A_{1}(x) \dots A_{n}(x): :B_{1}(y) \dots B_{m}(y): = 
:A_{1}(x) \dots A_{n}(x) B_{1}(y) \dots B_{m}(y): + 
\nonumber \\
\sum_{p=1}^{min(n,m)} \sum_{\rm partitions} \prod_{t=1}^{p} 
<\Phi_{0},A_{i_{t}}(x) B_{j_{t}}(y)\Phi_{0}>
:A_{i'_{1}}(x) \dots A_{i'_{n-p}}(x) B_{j'_{1}}(y) \dots B_{j'_{m-p}}(y): 
\end{eqnarray}
where 
$
\{i_{1},\dots,i_{p}\}, \quad \{i'_{1},\dots,i'_{n-p}\}
$
is a partition of
$1,\dots,n$,
and
$
\{j_{1},\dots,j_{p}\}, \quad \{j'_{1},\dots,j'_{m-p}\}
$
is a partition of
$1,\dots,m$;
the sum runs over all such partitions.

We conclude with a remark concerning the choice of the statistics. One can
also quantize the scalar field using Fermi statistics; in this way one will
loose only the causality property (\ref{causal}). However, such unphysical
fields, called {\it ghosts} will be essential as technical devices in the
description of zero-mass fields as we will see in the next Section. We have to
modify, however Wick theorem, in the sense that the sign of the permutation
\be
(1,2,\dots,n,1,2,\dots,m) \mapsto 
(i_{t},\dots,i_{p},i'_{1},\dots,i'_{n-p},
j_{1},\dots,j_{p},j'_{1},\dots,j'_{m-p})
\ee
should appear.
\newpage

\subsection{Perturbation Theory in the Causal Approach\label{pt}}

Perturbation theory, relies considerably on the axiom of causality, as shown 
by H. Epstein and V. Glaser \cite{EG1}. According to Bogoliubov and Shirkov, 
the $S$-matrix is constructed inductively order by order as a formal series
of operator valued distributions:
\be
S(g)=1+\sum_{n=1}^\infty{i^{n}\over n!}\int_{\R^{4n}} dx_{1}\cdots dx_{n}\,
T_{n}(x_{1},\cdots, x_{n}) g(x_{1})\cdots g(x_{n}),
\label{S}
\ee
where 
$g(x)$ 
is a tempered test function that switches the interaction and 
$T_{n}$
are operator-valued distributions acting in the Fock space of some collection
of free fields. For instance, in \cite{EG1} (see also \cite{Gl}) one considers
a real free scalar field, i.e. one considers the the Fock space is that one
defined in the preceding subsection. These operator-valued distributions,
which are called {\it chronological products} should verify some properties
which can be argued starting from {\it Bogoliubov axioms}.

$\bullet$
First, it is clear that we can consider them {\it completely symmetrical} in
all variables without loosing generality:
\be
T_{n}(x_{P(1)},\cdots x_{P(n)}) = T_{n}(x_{1},\cdots x_{n}), \quad \forall
P \in {\cal P}_{n}.
\label{sym}
\ee

$\bullet$
Next, we must have {\it Poincar\'e invariance}:
\be
{\cal U}_{a,\Lambda} T_{n}(x_{1},\cdots, x_{n}) {\cal U}^{-1}_{a,\Lambda} =
T_{n}(\Lambda\cdot x_{1}+a,\cdots, \Lambda\cdot x_{n}+a), \quad \forall 
\Lambda \in {\cal L}^{\uparrow}.
\label{invariance}
\ee
In particular, {\it translation invariance} is essential for implementing
Epstein-Glaser scheme of renormalisation. 

$\bullet$
The central axiom seems to be the requirement of {\it causality} which can be
written compactly as follows. Let us firstly introduce some standard notations.
Denote by
$
V^{+} \equiv \{x \in \R^{4} \vert \quad x^{2} > 0, \quad x_{0} > 0\}
$
and
$
V^{-} \equiv \{x \in \R^{4} \vert \quad x^{2} > 0, \quad x_{0} < 0\}
$
the upper (lower) lightcones and by 
$\overline{V^{\pm}}$
their closures. If
$X \equiv \{x_{1},\cdots, x_{m}\} \in \R^{4m}$
and
$Y \equiv \{y_{1},\cdots, y_{n}\} \in \R^{4m}$
are such that
$
x_{i} - y_{j} \not\in \overline{V^{-}}, \quad \forall i=1,\dots,m,\quad
j=1,\dots,n
$
we use the notation
$X \geq Y.$
We use the compact notation
$T_{n}(X) \equiv T_{n}(x_{1},\cdots, x_{n})$
and by 
$X \cup Y$
we mean the juxtaposition of the elements of $X$ and $Y$. In particular,
the expression
$T_{n+m}(X \cup Y)$
makes sense because of the symmetry property (\ref{sym}). Then the causality
axiom writes as follows:
\be
T_{n+m}(X \cup Y) = T_{m}(X) T_{n}(Y), \quad \forall X \geq Y.
\ee

$\bullet$
The {\it unitarity} of the $S$-matrix can be most easily expressed (see
\cite{EG1}) if one introduces, the following formal series:
\be
\bar{S}(g)=1+\sum_{n=1}^\infty{(-i)^{n}\over n!}\int_{\R^{4n}} 
dx_{1}\cdots dx_{n}\,\bar{T}_{n}(x_{1},\cdots, x_{n}) g(x_{1})\cdots g(x_{n}),
\label{barS}
\ee
where, by definition:
\be
(-1)^{|X|} \bar{T}_{n}(X) \equiv \sum_{r=1}^{n} (-1)^{r} 
\sum_{n_{1}+\cdots +n_{r}=n} \sum_{partitions} 
T_{n_{1}}(X_{1})\cdots T_{n_{r}}(X_{r});
\label{antichrono}
\ee
here 
$X_{1},\cdots,X_{r}$
is a partition of $X$, $|X|$ is the cardinal of the set $X$ and the sum runs
over all partitions. For instance, we have:
\be
\bar{T}_{1}(x) = T_{1}(x)
\ee
and
\be
\bar{T}_{2}(x,y) = - T_{2}(x,y) + T_{1}(x) T_{1}(y) + T_{1}(y) T_{1}(x).
\ee

One calls the operator-valued distributions 
$\bar{T_{n}}$
{\it anti-chronological products}. It is not very hard to prove that the series
(\ref{barS}) is the inverse of the series (\ref{S}) i.e. we have:
\be
\bar{S}(g) = S(g)^{-1}
\ee
as formal series. Then the unitarity axiom is:
\be
\bar{T}_{n}(X) = T_{n}(X)^{\dagger}, \quad \forall n \in \N, \quad \forall X.
\label{unitarity}
\ee

$\bullet$
The existence of the {\it adiabatic limit} can be formulated as follows. Let us
take in (\ref{S}) 
$g \rightarrow g_{\epsilon}$
where 
$\epsilon \in \R_{+}$
and
\be
g_{\epsilon}(x) \equiv g(\epsilon x).
\ee 

Then one requires that the limit
\be
S \equiv \lim_{\epsilon \searrow 0} S(g_{\epsilon})
\label{adiabatic}
\ee
exists, in the weak sense, and is independent of the the test function $g$. In
other words, the operator $S$ should depend only on the {\it coupling constant}
$g \equiv g(0).$ 
Equivalently, one requires that the limits
\be
T_{n} \equiv \lim_{\epsilon \searrow 0} T_{n}(g_{\epsilon}^{\otimes n}),
\quad n \geq 1
\ee
exists, in the weak sense, and are independent of the test function $g$. One 
also calls the limit performed above, the {\it infrared limit}. 

$\bullet$
Finally, one demands the {\it stability of the vacuum} i.e.
\be
\lim_{\epsilon \searrow 0} < \Phi_{0},  S(g_{\epsilon}) \Phi_{0}> = 1
\ee
or,
\be
\lim_{\epsilon \searrow 0} 
< \Phi_{0},  T_{n}(g^{\otimes n}_{\epsilon}) \Phi_{0}> = 0, 
\quad \forall n \in \N^{*}.
\ee

A {\it renormalisation theory} is the possibility to construct such a
$S$-matrix starting from the first order term:
\be
T_{1}(x) \equiv {\cal L}(x)
\ee
where  
${\cal L}$ 
is a Wick polynomial called {\it interaction Lagrangian} which should verify
the following axioms:
\be
{\cal U}_{a,\Lambda} {\cal L}(x) {\cal U}^{-1}_{a,\Lambda} =
{\cal L}(\Lambda\cdot x+a), 
\quad \forall \Lambda \in {\cal L}^{\uparrow},
\label{inv1}
\ee
\be
\left[{\cal L}(x), {\cal L}(y)\right] = 0, \quad \forall x,y \in \R^{4} \quad
s.t. \quad (x-y)^{2} < 0,
\label{causality1}
\ee
\be
{\cal L}(x)^{\dagger} = {\cal L}(x)
\label{unitarity1}
\ee
and
\be
L \equiv \lim_{\epsilon \searrow 0} {\cal L}(g_{\epsilon})
\label{adiabatic1}
\ee
should exists, in the weak sense, and should be independent of the test
function $g$. Moreover, we should have
\be
<\Phi_{0}, L \Phi_{0}> = 0.
\label{stability1}
\ee

In the analysis of Epstein and Glaser \cite{EG1}, \cite{Gl} it is proved that
if one considers a renormalisation theory for a real scalar free field of
positive mass i.e. one works in the corresponding Fock space (see the first
subsection) and takes ${\cal L}(x)$ to be a Wick polynomial, then such a
$S$-matrix exists. In the more familiar physicists language, {\it no
ultraviolet divergences} and {\it no infrared divergences} appear, i.e. the
$T_{n}$'s are finite and well defined and the adiabatic limit exists.  The only
remnant of the ordinary renormalisation theory is a non-uniqueness of the
$T_{n}$'s due to {\it finite} normalization terms which are distributions with
the support 
$\{x_{1} = \cdots = x_{n} = 0\}.$ 
The whole construction amounts to the operation of {\it distribution
splitting}. In this context appears what is sometimes called the 
{\it normalisation} problem. First, we define the {\it power} of a Wick
monomial $L$ as 
\be
\omega(L) \equiv n_{b} + 3/2~n_{f} + n_{d}
\ee
where 
$n_{b} (n_{f}) $ 
is the number of Bosonic (Fermionic) factors and
$n_{d}$
is the number of derivatives.  Suppose that the interaction Lagrangian is a
Wick polynomial of power $l$. Then the finite renormalisations of the type
described above are in fact Wick monomials. If we admit finite normalisation
which are producing Wick expressions of power strictly greater than $l$, then
this would amount to corrections to the interaction Lagrangian which have been
neglected from the very beginning. So, it is natural to impose that the
arbitrariness involved in the process of distribution splitting, i.e. the
finite normalisations should produce Wick monomials of power less or at most
equal to the power of the interaction Lagrangian. We call this assumption the
{\it normalisation axiom}.

For the case of the real scalar field one can prove that a renormalisation
theory do exist if the interaction Lagrangian verifies the properties 
(\ref{inv1}), (\ref{causality1}), (\ref{unitarity1}) and (\ref{adiabatic1}).
That's it, one can use the normalisation arbitrariness to construct
recurringly the chronological products verifying all the axioms. The main
advantage of Epstein-Glaser approach is a direct control of the fulfilment of
all axioms and the fact that one does not need a regularisation scheme.

It is instructive to show how the second
order distribution $T_{2}$ can be constructed. First one constructs the
distribution 
$D_{2}(x,y)$
\be
D_{2} (x,y) \equiv [T_{1}(x),T_{1}(y)].
\label{d2}
\ee
Form the axiom (\ref{causality1}) one has:
\be
{\rm supp}(D_{2}) = \{x-y \in V^{+} \cup V^{-}\}
\label{supp-d2}
\ee
i.e. the distribution 
$D_{2}$ 
has causal support. Then 
$D_{2}$ 
is split into a retarded and an advanced part 
\be
D_{2} = R_{2} - A_{2}
\label{split}
\ee
with
\be
{\rm supp}(R_{2}) = \{x-y \in V^{-} \}
\ee
and
\be
{\rm supp}(A_{2}) = \{x-y \in V^{+} \} \quad.
\ee 

Finally 
$T_{2}$ 
is given by
\be
T_{2}(x,y) = R_{2}(x,y) + T_{1}(y)T_{1}(x) = A_{2}(x,y) - T_{1}(x)T_{1}(y).
\label{t2}
\ee

This solution is arbitrary up to an operator-valued distribution with support
$\{x_{1} = x_{2}\}.$ 
One can use this arbitrariness to satisfy the axioms of relativistic
invariance and unitarity up to order 2 of the perturbation theory.

We mention now that if the interaction Lagrangian is a Wick polynomial, say
$:\varphi^{l}:(x)$,
then the induction procedure leads to the following generic expression for the
chronological products:
\be
T_{n}(g) = \int_{(\R^{4})^{\times n}} dx_{1}\cdots dx_{n} g(x_{1},\dots,x_{n})
\sum_{r_{1},\dots,r_{n} \leq l} F_{r_{1},\dots,r_{n}}(x_{1},\dots,x_{n})
:\varphi^{r_{1}}(x_{1})\cdots \varphi^{r_{n}}(x_{n}):
\label{chrono-n}
\ee
where the $F$'s are an ordinary distribution verifying translation invariance.
The existence of such expressions as well defined operators is a consequence of
the so-called {\it zero theorem} \cite{EG1}.

We close this Section with a comment about the possible choices of the
interaction Lagrangian. Let us suppose that the interaction Lagrangian is of 
the form
$:\varphi^{l}:$.
with $l = 2$; then one can prove that the matrix element
$<p_{1},\dots,p_{n}|\int_{\R^{4}} dx :\varphi^{2}:(x) | q_{1},\dots,q_{n}>$
is a sum of the type
$\delta(p_{1}-q_{i_{1}}) \cdots \delta(p_{n}-q_{i_{n}})$
which describes no interaction; a similar result is valid for the case $l = 1$.
\newpage

\section{The Quantisation of the Electromagnetic Field}
\subsection{Quantisation without Ghost Fields\label{quant}}

We remind briefly the analysis from \cite{WG}, \cite{SW1}, \cite{Ka}. The idea
is to apply the prescription from subsection \ref{sq} to the Hilbert space of
the photon 
${\rm H}_{photon}$
given by (\ref{photon}). The idea is to express the (Bosonic) {\it Fock space
of the photon} 
\be
{\cal F}_{photon} \equiv {\cal F}^{+}({\rm H}_{photon}) 
\ee
as a factorization of the type (\ref{photon}). It is natural to start with the
``bigger"  Fock space
\be
{\cal H} \equiv {\cal F}^{+}({\rm H}) \equiv \oplus_{n\geq 0} {\cal H}_{n}.
\label{aux-ph}
\ee

The first observation is that one can canonically identify the
$n^{th}$-particle subspace
${\cal H}_{n}$
with the set of Borel functions
$\Phi^{(n)}_{\mu_{1},\dots,\mu_{n}}: (X^{+}_{0})^{\times n} \rightarrow \C$
which are square summable:
\be
\int_{(X^{+}_{0})^{\times n}} \prod_{i=1}^{n} d\alpha^{+}_{0}(k_{i})
\sum_{\mu_{1},\dots,\mu_{n} =0}^{3}
|\Phi^{(n)}_{\mu_{1},\dots,\mu_{n}}(k_{1},\dots,k_{n})|^{2} < \infty
\ee
and verify the symmetry property
\be
\Phi^{(n)}_{\mu_{P(1)},\dots,\mu_{P(n)}}(k_{P(1)},\dots,k_{P(n)}) =
\Phi^{(n)}_{\mu_{1},\dots,\mu_{n}}(k_{1},\dots,k_{n}), 
\quad \forall P \in {\cal P}_{n}.
\label{symmetry1}
\ee

In ${\cal H}$ the expression of the scalar product is:
\be
<\Psi,\Phi> \equiv \overline{\Psi^{(0)}} \Phi^{(0)} + \sum_{n=1}^{\infty}
\int_{(X^{+}_{0})^{\times n}} \prod_{i=1}^{n} d\alpha^{+}_{0}(k_{i})
\sum_{\mu_{1},\dots,\mu_{n} =0}^{3}
\overline{\Psi^{(n)}_{\mu_{1},\dots,\mu_{n}}(k_{1},\dots,k_{n})}
\Phi^{(n)}_{\mu_{1},\dots,\mu_{n}}(k_{1},\dots,k_{n})
\label{scalar-prod1}
\ee
and we have a (non-unitary) representation of the Poincar\'e group given by:
\be
{\cal U}_{g} \equiv \Gamma(U_{g}), \quad \forall g \in {\cal P};
\label{reps1}
\ee
here $U_{g}$ is given by (\ref{reps0}).

Let us define the following {\it Krein operator}
\be
G \equiv \Gamma(-g) = \sum_{n\geq 0} (-g)^{\otimes n}
\label{Krein-em}
\ee
where the operator $g$ appears in (\ref{g}). Then we can define the following
non-degenerate sesquilinear form on ${\cal H}$:
\be
(\Psi,\Phi) \equiv <\Psi, G \Phi>
\ee
or, explicitly:
\be
(\Psi,\Phi) \equiv \overline{\Psi^{(0)}} \Phi^{(0)} + \sum_{n=1}^{\infty}
(-1)^{n} \int_{(X^{+}_{0})^{\times n}} \prod_{i=1}^{n} 
\left[ d\alpha^{+}_{0}(k_{i}) g^{\mu_{i}\nu_{i}} \right]
\overline{\Psi^{(n)}_{\mu_{1},\dots,\mu_{n}}(k_{1},\dots,k_{n})}
\Phi^{(n)}_{\nu_{1},\dots,\nu_{n}}(k_{1},\dots,k_{n}).
\label{sesq1}
\ee

Then the sesquilinear form 
$(\cdot,\cdot)$
behaves naturally with respect to the action of the Poincar\'e group:
\be
({\cal U}_{g}\Psi,{\cal U}_{g}\Phi) = (\Psi,\Phi), 
\quad \forall g \in {\cal P}^{\uparrow}, \quad 
({\cal U}_{I_{t}}\Psi,{\cal U}_{I_{t}}\Phi) = \overline{(\Psi,\Phi)}.
\ee

We denote 
$|\phi|^{2} = <\phi,\phi>$
and
$\Vert\Phi\Vert^{2} = (\Phi,\Phi)$.

Now one has from lemma \ref{h'}:
\begin{lemma}
Let us consider the following subspace of ${\cal H}$:
\be
{\cal H}' \equiv {\cal F}^{+}({\rm H}') = \oplus_{n\geq 0} {\cal H}_{n}'.
\ee

Then
${\cal H}_{n}', \quad n \geq 1$
is generated by elements of the form
$
\phi_{1} \vee \cdots \vee \phi_{n}, \quad 
\phi_{1},\dots,\phi_{n} \in {\rm H}'
$
and, in the representation adopted previously for the Hilbert space
${\cal H}_{n}$
we can take
\be
{\cal H}_{n}' = \{ \Phi^{(n)} \in {\cal H}_{n} | \quad
k_{1}^{\nu_{1}} \Phi^{(n)}_{\nu_{1},\dots,\nu_{n}}(k_{1},\dots,k_{n}) = 0\}.
\label{h-prim}
\ee

Moreover, the sesquilinear form
$\left. (\cdot,\cdot)\right|_{{\cal H}'}$
is positively defined.
\label{h'1}
\end{lemma}

Next, one has the analogue of lemma \ref{h''}:
\begin{lemma}

Let ${\cal H}'' \subset {\cal H}'$
given by
\be
{\cal H}'' \equiv \{\Phi \in {\cal H}' | \quad \Vert \Phi \Vert^{2} = 0\} =
\oplus_{n\geq 0} {\cal H}''_{n}
\ee

Then, the subspace
${\cal H}''_{n}, \quad n \geq 1$
is generated by elements of the type
$
\phi_{1} \vee \cdots \vee \phi_{n}
$
where at least one of the vectors
$\phi_{1},\dots,\phi_{n} \in {\rm H}'$
belongs to
${\rm H}''$.

Moreover, in the representation adopted previously for the Hilbert space
${\cal H}_{n}$
the elements of
${\cal H}''_{n}$
are linearly generated by functions of the type:
\be
\Phi^{(n)}_{\nu_{1},\dots,\nu_{n}}(k_{1},\dots,k_{n}) =
{1\over n} \sum_{i=1}^{n} (k_{i})_{\nu_{i}} \lambda(k_{i})
\Psi^{(n-1)}_{\nu_{1},\dots,\hat{\nu_{i}},\dots,\nu_{n}}
(k_{1},\dots,\hat{k_{i}},\dots,k_{n})
\label{h-secund}
\ee
with
$\Psi \in {\cal H}'$
and
$\lambda: X^{+}_{0} \rightarrow \C$
arbitrary.
\label{h''2}
\end{lemma}

Finally we have:
\begin{prop}
There exists an canonical isomorphism of Hilbert spaces
\be
{\cal F}_{photon} \simeq \overline{{\cal H}'/{\cal H}''}.
\ee
\label{photon-factor}
\end{prop}

{\bf Proof:}
If 
$\psi \in {\rm H}'$
then we denote its class with respect to
${\rm H}''$
by
$[\psi]$;
similarly, if
$\Phi \in {\cal H}'$
we denote its class with respect to
${\cal H}''$
by
$[\Phi].$
Then the application
$
B: \overline{{\cal H}'/{\cal H}''} \rightarrow {\cal F}_{photon}
$ 
is well defined by linearity, continuity and
\be
B([\phi_{1} \vee \cdots \vee \phi_{n}]) \equiv 
[\phi_{1}] \vee \cdots \vee [\phi_{n}]
\ee
and it is the desired isomorphism. Moreover, the sesquilinear form
$(\cdot,\cdot)$
is strictly positive defined on the factor space, so it induces a scalar
product.
$\qed$

Now we can define the electromagnetic field as an operator on the Hilbert space
${\cal H}$
in analogy to the construction from subsections \ref{sq} and \ref{ff}
(see the relations (\ref{annihilation}) and (\ref{creation})): 
we define for every
$p \in X^{+}_{0}$
the annihilation and creation operators 

\be
\left( A_{\nu}(p) \Phi\right)^{(n)}_{\mu_{1},\dots,\mu_{n}}
(k_{1},\dots,k_{n}) \equiv \sqrt{n+1}
\Phi^{(n+1)}_{\nu,\mu_{1},\dots,\mu_{n}}(p,k_{1},\dots,k_{n})
\label{annihilation-em}
\ee
and
\be
\left( A^{\dagger}_{\nu}(p) \Phi\right)^{(n)}_{\mu_{1},\dots,\mu_{n}}
(k_{1},\dots,k_{n}) \equiv - 2 \omega({\bf p}) {1\over \sqrt{n}} 
\sum_{i=1}^{n} \delta({\bf p}-{\bf k_{i}}) g_{\nu\mu_{i}} 
\Phi^{(n-1)}_{\mu_{1},\dots,\hat{\mu_{i}},\dots,\mu_{n}}
(k_{1},\dots,\hat{k_{i}},\dots,k_{n}).
\label{creation-em}
\ee

Then one has the {\it canonical commutation relations} (CAR)
\be
\left[ A_{\nu}(p), A_{\rho}(p')\right] = 0,\quad
\left[ A^{\dagger}_{\nu}(p), A^{\dagger}_{\rho}(p')\right] = 0, \quad
\left[ A_{\nu}(p), A^{\dagger}_{\rho}(p')\right] = 
- 2 \omega({\bf p}) g_{\nu\rho} \delta{({\bf p}-{\bf p}')} {\bf 1}
\label{CAR-em}
\ee
and the relation
\be
(A^{\dagger}_{\nu}(p)\Psi,\Phi) = (\Psi,A_{\nu}(p)\Phi), \quad \forall 
\Psi, \Phi \in {\cal H}
\ee
which shows that
$A_{\nu}^{\dagger}(p)$
is the adjoint of 
$A_{\nu}(p)$
with respect to the sesquilinear form
$(\cdot,\cdot)$.

We also have a natural behaviour with respect to the action of the Poincar\'e
group (see (\ref{repsA})):
\be
{\cal U}_{a,\Lambda} A_{\nu}(p) {\cal U}^{-1}_{a,\Lambda}  = 
e^{ia\cdot p} (\Lambda^{-1})_{\nu}^{\ \rho} A_{\rho}(\Lambda \cdot p),
\quad \forall \Lambda \in {\cal L}^{\uparrow}, \quad
{\cal U}_{I_{t}} A_{\nu}(p) {\cal U}^{-1}_{I_{t}}  = 
(I_{t})_{\nu}^{\ \rho} A_{\rho}(I_{s} \cdot p)
\label{repsA-em}
\ee
and a similar relation for 
$A_{\nu}^{\dagger}(p)$.

Now we define the {\it electromagnetic field in the point x} according to
\be
A_{\nu}(x) \equiv A^{(+)}_{\nu}(x) + A^{(-)}_{\nu}(x)
\label{em}
\ee
where the expressions appearing in the right hand side are the positive
(negative) frequency parts and are defined by:
\be
A^{(+)}_{\nu}(x) \equiv {1\over (2\pi)^{3/2}} \int_{X^{+}_{0}}
d\alpha^{+}_{0}(p) e^{ip\cdot x} A^{\dagger}_{\nu}(p),\quad
A^{(-)}_{\nu}(x) \equiv {1\over (2\pi)^{3/2}} \int_{X^{+}_{0}}
d\alpha^{+}_{0}(p) e^{-ip\cdot x} A_{\nu}(p).
\label{em-pm}
\ee

The explicit expressions are
\be
\left(A^{(+)}_{\nu}(x) \Phi\right)^{(n)}_{\mu_{1},\dots,\mu_{n}}
(k_{1},\dots,k_{n}) = {\sqrt{n+1} \over (2\pi)^{3/2}} \int_{X^{+}_{0}}
d\alpha^{+}_{0}(p) e^{ip\cdot x} \Phi^{(n+1)}_{\nu,\mu_{1},\dots,\mu_{n}}
(p,k_{1},\dots,k_{n})
\ee
and
\be
\left(A^{(-)}_{\nu}(x) \Phi\right)^{(n)}_{\mu_{1},\dots,\mu_{n}}
(k_{1},\dots,k_{n}) = - {1\over (2\pi)^{3/2}\sqrt{n}} \sum_{i=1}^{n}
e^{i k_{i}\cdot x} g_{\nu\mu_{i}} 
\Phi^{(n-1)}_{\mu_{1},\dots,\hat{\mu_{i}},\dots,\mu_{n}}
(k_{1},\dots,\hat{k_{i}},\dots,k_{n}).
\ee

If 
$f \in {\cal S}(\R^{4},\R^{4})$
we have the well-defined operators
\be
A(f) \equiv \int_{\R^{4}} dx f^{\mu}(x) A_{\mu}(x)
\label{Af}
\ee
or, more explicitly:
\be
A(f) = \sqrt{2\pi} \int_{X^{+}_{0}} d\alpha^{+}_{0} 
\left[ \tilde{f^{\mu}}(p) A_{\mu}(p) + 
\tilde{f^{\mu}}(-p) A^{\dagger}_{\mu}(p)\right].
\label{Aff}
\ee

The properties of the electromagnetic field operator 
$A_{\nu}(x)$
are summarised in the following elementary proposition:
\begin{prop}
The following relations are true:
\be
(A_{\nu}(x)\Psi,\Phi) = (\Psi,A_{\nu}(x)\Phi), \quad \forall 
\Psi, \Phi \in {\cal H},
\label{AA-dagger}
\ee
\be
\square A_{\nu}(x) = 0
\label{eqA}
\ee
and
\be
\left[A^{(\mp)}_{\mu}(x),A^{(\pm)}_{\nu}(y)\right] = 
- g_{\mu\nu} D^{(\pm)}_{0}(x-y) \times {\bf 1},\quad
\left[A^{(\pm)}_{\mu}(x),A^{(\pm)}_{\nu}(y)\right] = 0.
\label{CCR-em-pm}
\ee
As a consequence we also have:
\be
\left[A_{\mu}(x),A_{\nu}(y)\right] = - g_{\mu\nu} D_{0}(x-y) \times {\bf 1};
\label{commutation-em}
\ee
here
\be
D_{0}(x) = D_{0}^{(+}(x) + D_{0}^{(-)}(x)
\ee
is the Pauli-Jordan distribution and
$D^{(\pm)}(x)$
are given by:
\be
D_{0}^{(\pm)}(x) \equiv \pm {1\over (2\pi)^{3/2}} \int_{X^{+}_{0}}
d\alpha^{+}_{0}(p) e^{\mp i p\cdot x}.
\label{D0}
\ee
\end{prop}

Let us note that we have:
\be
\square D_{0}^{(\pm)}(x) = 0, \quad \square D_{0}(x) = 0.
\ee

One can describe in very convenient way the subspaces
${\cal H}'$
and
${\cal H}''$
using the following operators
\be
L(x) \equiv \partial^{\mu} A^{(-)}_{\mu}(x), \quad
L^{\dagger}(x) \equiv \partial^{\mu} A^{(+)}_{\mu}(x).
\ee

Indeed, one has the following result:
\begin{prop}
The following relations are true:
\be
{\cal H}' = \{ \Phi \in {\cal H} | \quad L(x) \Phi = 0, 
\quad \forall x \in \R^{4} \} = \cap_{x \in \R^{4}} Ker(L(x))
\ee
and
\be
{\cal H}'' = \{ L(x)^{\dagger}\Phi | \quad \forall \Phi \in {\cal H}, \quad
\quad \forall x \in \R^{4} \} = \cup_{x \in \R^{4}} Im(L(x)^{\dagger}).
\ee

It follows that we have
\be
{\cal F}_{photon} = \overline{\cap_{x \in \R^{4}} Ker(L(x))/
\cup_{x \in \R^{4}} Im(L(x)^{\dagger})}.
\ee
\label{photon-fock}
\end{prop}

Let us note that we have:
\be
[L(x), L^{\dagger}(y)] = 0, \quad \forall x,y \in \R^{4}; \quad
[L(f), L^{\dagger}(g)] = 0, \quad \forall f, g \in {\cal S}(\R^{4}).
\label{LL-dagger}
\ee

We concentrate now on the constructions of observables on the Fock space of the
photon
${\cal F}_{photon}$.
Because this space is rather hard to manipulate, we will distinguish a class of
observables which are induced by self-adjoint operators on the Hilbert space 
${\cal H}$. 
Indeed, if $O$ is such an operator and it leaves invariant the subspaces 
${\cal H}'$
and
${\cal H}''$
then it factorizes to an operator on
${\cal F}_{photon}$
according to the formula
\be
[O] [\Phi] \equiv [O\Phi].
\label{O-inv}
\ee

This type of observables are called {\it gauge invariant observables}. It is
clear that not all observables on
${\cal F}_{photon}$
are of this type.  

Now we have the following result:
\begin{lemma}
An operator
$O: {\cal H} \rightarrow {\cal H}$
induces a gauge invariant observable if and only if it verifies:
\be
\left. [L(x), O]\right|_{{\cal H}'} = 0, \quad \forall x \in \R^{4}.
\label{gauge-inv}
\ee
\label{gi}
\end{lemma}

{\bf Proof:}
If the relation (\ref{gauge-inv}) is true, then one applies the proposition
\ref{photon-fock} and it is clear that the operator $O$ leaves the subspaces 
${\cal H}'$
and
${\cal H}''$
invariant, so it induces the operator
$[O]$.
Conversely, if the operator $O$ induces an operator 
$[O]$
then it should leave invariant the subspace
${\cal H}'$ 
i.e. if
$\Phi \in {\cal H}'$
then we should also have
$O\Phi \in {\cal H}'$.
Equivalently, we use proposition \ref{photon-fock} and obtain that if
$L(x) \Phi = 0, \quad \forall x \in \R^{4}$
then
$L(x) O\Phi = 0, \quad \forall x \in \R^{4}$.
But this implies that the relation (\ref{gauge-inv}) is true.
$\qed$

\begin{rem}
It is clear that the same result is true if one replaces in (\ref{gauge-inv})
the commutator with the anticommutator.
\end{rem}

In the end of this subsection we construct some typical gauge invariant
observables. The verification of the condition from the preceding lemma is
trivial. The first one is the so-called {\it strength of the
electromagnetic field} defined by:
\be
F_{\mu\nu} \equiv \partial_{\mu} A_{\nu}(x) - \partial_{\nu} A_{\mu}(x).
\label{F-mu-nu}
\ee

The second one is more subtle. 
\begin{prop}
Let us consider a subset of the test functions space:
\be
{\cal S}_{div} \equiv \{ f \in {\cal S}(\R^{4},\R^{4}) |\quad 
\partial_{\mu} f^{\mu} = 0\}.
\ee
For any 
$f \in {\cal S}_{div}$
the operator 
$A(f)$
(see (\ref{Af}) for the definition) induces a gauge invariant observable
$[A(f)]$.
Moreover, if we have
$f_{\mu} = \partial_{\mu} g$
for some test function verifying the equation
$\square g = 0$
then 
$[A(f)] = 0$.
\end{prop}

Now we can define, in analogy with the remark \ref{creation+annihilation},
annihilation and creation operators for the electromagnetic potentials as maps 
from
${\sf H}_{photon}$
into 
${\cal F}_{photon}$. 
We have:
\begin{prop}
Let
$f \in {\sf H}'$
and
$[f]$
its equivalence class modulo
${\sf H}''$.
Then the operators
$A^{\sharp}([f]): {\cal F}_{photon} \rightarrow {\cal F}_{photon}$
given by
\be
A([f]) [\Phi] = \left[ \int_{X^{+}_{0}} d\alpha^{+}_{0} 
\overline{f^{\mu}(p)} A_{\mu}(p) \Phi \right], \quad
A^{\dagger}([f]) [\Phi] = \left[ \int_{X^{+}_{0}} d\alpha^{+}_{0} 
f^{\mu}(p) A^{\dagger}_{\mu}(p) \Phi \right]
\ee
are well defined and one has the following formula
\be
[A(f)] = \sqrt{2\pi} \left( A([\overline{f_{+}}]) + 
A^{\dagger}([f_{-}])\right).
\label{em-potentials}
\ee
\end{prop}

Let us note that the expression (\ref{em-potentials}) is the perfect analogue
of the generic formula (\ref{free-scalar1}). However, a major difference
appears, namely one is not able to find analogues of the Wick monomials
inducing well defined expressions on
${\cal F}_{photon}$.
More precisely, one can define without any problems expressions of the type
\be
{\cal L}(g) = \int_{\R^{4}} dx g(x) \sum
j^{(\nu_{1},I_{1}),\dots,(\nu_{r},I_{r})}(x) 
:\partial_{I_{1}} A_{\nu_{1}}(x) \cdots \partial_{I_{r}} A_{\nu_{r}}(x):
\label{interaction}
\ee
for $r > 1$ on the Hilbert space ${\cal H}$, but it will be impossible to
factorize such an expression to 
${\cal F}_{photon}$
even if one considers it in the adiabatic limit.

We continue this subsection with the analysis of possible interactions between
the electromagnetic field and matter. Let us consider that the (Fock) space of
the ``matter" fields is denoted by 
${\cal H}_{matter}$. 
Then, in the hypothesis of weak coupling, one can argue that the Hilbert space
of the combined system photons $+$ matter is 
$ 
{\cal H}_{total} \equiv {\cal F}_{photon} \otimes {\cal H}_{matter}.  
$ 
It is easy to see that, if we define, 
$
\tilde{\cal H} \equiv {\cal H} \otimes {\cal H}_{matter}, \quad 
\tilde{\cal H}' \equiv {\cal H}' \otimes {\cal H}_{matter}
$
and
$
\tilde{\cal H}'' \equiv {\cal H}'' \otimes {\cal H}_{matter}
$
we have as before:
\be
{\cal H}_{total} \simeq \overline{\tilde{\cal H}'/ \tilde{\cal H}''}.
\ee

In the Hilbert space
$\tilde{\cal H}$
we can define as usual the expressions for the electromagnetic potentials and
all properties listed previously stay true. In particular, there are no
interactions of the type (\ref{interaction}). So, we must try to construct the
interaction in the form
\be
T_{1}(g) \equiv \int_{\R^{4}} dx g(x) A_{\nu}(x) j^{\nu}(x)
\label{inter}
\ee
where
$j^{\nu}(x)$
are some Wick polynomials in the ``matter" fields called {\it currents}. 
\begin{rem}
We note that it not necessary to consider more general expressions of the type
${\cal L}(x) = \sum j^{\nu,I}(x) \partial_{I}A_{\nu}(x)$
because, in the adiabatic limit, by partial integration, one can exhibit the 
integrated expression
$T_{1}(g) = \int_{\R^{4}} dx g(x) {\cal L}(x)$
into the form (\ref{inter}) from above.
\end{rem}

Then we have a simple but very important result:
\begin{thm}
The expression (\ref{inter}) induces, in the adiabatic limit, a well defined
expression on the Hilbert space
${\cal H}_{total}$
if and only if the current is conserved, i.e.
\be
\partial_{\mu}j^{\mu} = 0.
\label{conservation-1}
\ee
\end{thm}
{\bf Proof:}
One applies the criterion from lemma \ref{gi}.
$\qed$

Thus, we see that if we have a chiral interaction described by the axial
current 
$
j^{\mu}_{A}(x) = \overline{\psi}(x) \gamma^{\mu}\gamma_{5} \psi(x)
$
for some Dirac field of mass 
$m > 0$,
then, because as it is well known,
$
\partial_{\mu} j^{\mu}_{A}(x) \not= 0
$
this type of interaction does not produce a legitimate expression on the
physical Hilbert space
${\cal H}_{total}$.

This result admits a generalisation. In studying higher orders of the
perturbation theory, starting from an interaction Lagrangian of the type
(\ref{inter}), one gets for the chronological products, expressions of the type
(see (\ref{chrono-n})):
\begin{eqnarray}
T_{n}(g) = \int_{(\R^{4})^{\times n}} dx_{1}\cdots dx_{n} g(x_{1},\dots,x_{n})
\nonumber \\
\sum_{l=0}^{n} {1\over l!} \sum_{k_{1},\dots,k_{l}}
j^{\mu_{1},\dots,\mu_{l}}_{k_{1},\dots,k_{l}}(x_{1},\dots,x_{n})
:A_{\mu_{1}}(x_{k_{1}}) \cdots A_{\mu_{l}}(x_{k_{l}}):
\label{chrono-n-em}
\end{eqnarray}
where
$j^{\mu_{1},\dots,\mu_{l}}_{k_{1},\dots,k_{l}}(x_{1},\dots,x_{n})$
are some distribution-valued operators build form the matter fields called
{\it multi-currents}. The existence of such objects in the Hilbert space 
$\tilde{\cal H}$
is guaranteed by the {\it zero theorem} but we also have (see \cite{Sc1},
formula (4.6.9)):
\begin{thm}
The expressions (\ref{chrono-n-em}) are inducing well defined
expression on
${\cal H}_{total}$,
in the adiabatic limit, if and only if the following relations are verified:
\be
{\partial \over \partial x^{\nu_{1}}_{k_{1}}} 
j^{\mu_{1},\dots,\mu_{l}}_{k_{1},\dots,k_{l}}(x_{1},\dots,x_{n}) = 0
\label{conservation-n}
\ee
i.e. the multi-currents are also conserved. 
\end{thm}

Let us define now the so-called {\it gauge transformation} i.e. the operatorial
transformation
\be
\delta_{\xi} A_{\nu}(x) \equiv \partial_{\nu} \xi(x) \times {\bf 1}
\ee
where $\xi$ is a test function. Then one can easily see that the identities
(\ref{conservation-1}) and (\ref{conservation-n}) (see \cite{Sc1}) are
equivalent to the following equations expressing the gauge invariance of the
$S$-matrix:
\be
\lim_{\epsilon \searrow 0} \delta_{\xi} T_{n}(g_{\epsilon}^{\otimes n}) = 0,
\quad \forall n \in \N^{*} \quad \Leftrightarrow \quad 
\lim_{\epsilon \searrow 0} \delta_{\xi} S(g_{\epsilon}) = 0.
\ee

So we see that the gauge invariance of the $S$-matrix should not be considered
as an independent axiom, as it is usually done in the literature, but is a
consequence of the requirement that the $S$-matrix factorizes, in the adiabatic
limit, to the ``physical" Hilbert space
${\cal H}_{total}$.

We close this subsection with some comments concerning other conditions which
should be imposed to the currents. One should translate the requirements 
(\ref{inv1}), (\ref{causality1}) and (\ref{unitarity1}) and obtains that
the current 
$j^{\mu}(x)$
should be Poincar\'e covariant, should commute for spatially separated points
and should be hermitian.

\subsection{Quantisation with Ghost Fields\label{gh}}

In this subsection we follow the analysis of the ghosts fields from \cite{Kr2}
and then we show explicitly that one can obtain a new realisation of 
${\cal F}_{photon}$
which is essential for the construction of non-Abelian gauge fields. The
details of this analysis seems to be missing from the literature (although some
illuminating remarks do appear in \cite{We}).

We consider the Hilbert space
${\sf H}^{gh} \equiv L^{2}(X^{+}_{m},\C^{2},d\alpha_{m}^{+})$
with the scalar product:
\be
<\phi,\psi> \equiv \int_{X^{+}_{m}} d\alpha_{m}^{+} 
\left. <\phi(p), \psi(p)>\right|_{\C^{2}}.
\ee

In this space acts the following unitary representation of the Poincar\'e
group:
\be
\left( U_{a,\Lambda} \phi\right)(p) \equiv 
e^{ia\cdot p} \phi(\Lambda^{-1}\cdot p)\quad {\rm for} \quad \Lambda \in 
{\cal L}^{\uparrow}, \qquad 
\left( U_{I_{t}} \phi\right)(p) \equiv \overline{ \phi(I_{s}\cdot p)}.
\ee

The Fock space
${\cal F}^{-}({\sf H}^{gh})$
is called {\it ghost particle Hilbert space}. Remark the choice of the
Fermi-Dirac statistics which seems to be essential for the whole analysis. 

As in the subsection \ref{ff} one can canonically identify the 
$n^{\rm th}$-particle subspace
$
{\cal F}^{-}({\sf H}^{gh})
$
with the set of Borel functions
$\Phi^{(n)}: (X^{+}_{m})^{\times n} \rightarrow \C^{2}$
which are square integrable with respect to the product measure
$(\alpha_{m}^{+})^{\times n}$
and verify the symmetry property
\be
\Phi^{(n)}_{i_{P(1)},\dots,i_{P(n)}}(p_{P(1)},\dots,p_{P(n)}) = (-1)^{|P|}
\Phi^{(n)}_{i_{1},\dots,i_{n}}(p_{1},\dots,p_{n}), 
\quad \forall P \in {\cal P}_{n}.
\label{sym-ghost}
\ee

On the dense domain of test functions one can define the annihilation
and creation operators for any
$j = 1, 2$: 
\be
\left( d_{j}(k) \Phi\right)^{(n)}_{i_{1},\dots,i_{n}}(p_{1},\dots,p_{n}) 
\equiv \sqrt{n+1} \Phi^{(n+1)}_{j,i_{1},\dots,i_{n}}(k,p_{1},\dots,p_{n})
\label{annihilation-gh}
\ee
and
\be
\left( d_{j}^{*}(k) \Phi\right)^{(n)}_{i_{1},\dots,i_{n}}
(p_{1},\dots,p_{n}) \equiv 
2 \omega({\bf k}) {1\over \sqrt{n}} \sum_{s=1}^{n} (-1)^{s-1}
\delta({\bf k}-{\bf p_{s}}) \delta_{ji_{s}}
\Phi^{(n-1)}_{i_{1},\dots,\hat{i_{s}},\dots,i_{n}}
(p_{1},\dots,\hat{p_{s}},\dots,p_{n}).
\label{creation-gh}
\ee

They verify canonical anticommutation relations
\be
\{d_{j}(k),d_{k}(q)\} = 0, \quad 
\{d^{*}_{j}(k),d^{*}_{k}(q)\} = 0, \quad 
\{d_{j}(k),d^{*}_{k}(q)\} = \delta_{jk} 2 \omega({\bf q}) 
\delta({\bf k} - {\bf q}) {\bf 1}
\ee
and behave naturally with respect to Poincar\'e transform (see (\ref{repsA})).

It is convenient to denote
\be
b^{\sharp}(p) \equiv d^{\sharp}_{1}(p), 
\quad c^{\sharp}(p) \equiv d^{\sharp}_{2}(p).
\ee

Then the Fermionic fields
\be
u(x) \equiv {1\over (2\pi)^{3/2}} \int_{X^{+}_{0}} d\alpha^{+}_{0}(q)
\left[ e^{-i q\cdot x} b(q) + e^{i q\cdot x} c^{*}(q) \right]
\label{u}
\ee
and
\be
\tilde{u}(x) \equiv {1\over (2\pi)^{3/2}} \int_{X^{+}_{0}} d\alpha^{+}_{0}(q)
\left[ - e^{-i q\cdot x} c(q) + e^{i q\cdot x} b^{*}(q) \right]
\label{u-tilde}
\ee
are called {\it ghost fields}. They verify the wave equations:
\be
\square u(x) = 0, \quad \square \tilde{u}(x) = 0
\label{equ}
\ee
and if we identify, as usual the positive (negative) frequency parts we have
the canonical anticommutation relations:
\begin{eqnarray}
\{ u^{(\epsilon)}(x),u^{(\epsilon')}(y)\} = 0, \quad
\{u(x),u(y)\} = 0, \quad
\{ \tilde{u}^{(\epsilon)}(x),\tilde{u}^{(\epsilon')}(y)\} = 0, \quad
\{\tilde{u}(x),\tilde{u}(y)\} = 0, 
\nonumber \\
\{ u^{(\epsilon)}(x),\tilde{u}^{(-\epsilon)}(y)\} = D^{(-\epsilon)}_{0}(x-y), 
\quad \{u(x),\tilde{u}(y)\} = D_{0}(x-y) 
\quad \forall \epsilon, \epsilon' = \pm.
\label{CAR-ghosts}
\end{eqnarray}

Now we consider that we are working into the Fock space 
\be
{\cal H}^{gh} \equiv {\cal H} \otimes {\cal F}^{-}({\sf H}^{gh})
\ee
i.e. we tensor the auxiliary Fock space 
${\cal H}$ 
used in the construction of the photon Fock space with the ghost Fock space
(see (\ref{aux-ph})). In this Hilbert space, we can define without problems the
electromagnetic potential and the ghosts.  Now we can introduce an important
operator:
\be
Q \equiv \int_{X^{+}_{0}} d\alpha^{+}_{0}(q) k^{\mu}
\left[ A_{\mu}(k) c^{*}(k) + A^{\dagger}_{\mu}(k) b(k)\right]
\label{supercharge}
\ee
called {\it supercharge}. Its properties are summarised in the following
proposition which can be proved by elementary computations:
\begin{prop}
The following relations are valid:
\be
Q \Phi_{0} = 0;
\label{Q-0}
\ee
\be
\left[ Q, A^{\dagger}_{\mu}(k) \right] = k_{\mu} c^{*}(k),\quad
\left\{ Q, b^{*}(k) \right\} = k^{\mu} A^{\dagger}_{\mu}(k), \quad
\left\{ Q, c^{*}(k) \right\} = 0;
\label{com-dagger}
\ee
\be
\left[ Q, A_{\mu}(k) \right] = k_{\mu} b(k),\quad
\left\{ Q, b(k) \right\} = 0, \quad
\left\{ Q, c(k) \right\} = k^{\mu} A_{\mu}(k);
\label{com}
\ee
\be
Q^{2} = 0;
\label{square}
\ee
\be
Im(Q) \subset Ker(Q)
\label{im-ker}
\ee
and 
\be
{\cal U}_{g} Q = Q {\cal U}_{g}, \quad \forall g \in {\cal P}.
\label{UQ}
\ee

Moreover, one can express the supercharge in terms of the ghosts fields as
follows:
\be
Q = \int_{\R^{3}} d^{3}x  \partial^{\mu} A_{\mu}(x) 
\stackrel{\leftrightarrow}{\partial_{0}}u(x).
\ee
\end{prop}

In particular (\ref{square}) justify the terminology of supercharge and
(\ref{im-ker}) indicates that it might be interesting to take the quotient.
Indeed, we will rigorously prove that this quotient coincides with 
${\cal F}_{photon}$.

First, we give a more convenient representation for the Hilbert space
${\cal H}^{gh}$, namely
\be
{\cal H}^{gh} = \sum_{n,m,l=0}^{\infty} {\cal H}_{nml}
\ee
where
${\cal H}_{nml}$
consists of Borel functions
$
\Phi^{(nml)}_{\mu_{1},\dots,\mu_{n}}: (X^{+}_{0})^{n+m+l} \rightarrow \C
$
such that
\be
\sum_{n,m,l=0}^{\infty} \int_{(X^{+}_{0})^{n+m+l}} d\alpha^{+}_{0}(K)
d\alpha^{+}_{0}(P) d\alpha^{+}_{0}(Q) 
\sum_{\mu_{1},\dots,\mu_{n}=0}^{3} 
|\Phi^{(nml)}_{\mu_{1},\dots,\mu_{n}}(K,P,Q)| \leq \infty
\ee
(here 
$
K \equiv (k_{1},\dots,k_{n}), \quad
P \equiv (p_{1},\dots,p_{m})
$
and
$
Q \equiv (q_{1},\dots,q_{l}))
$
and verify the symmetry property
\begin{eqnarray}
\Phi^{(nml)}_{\mu_{P(1)},\dots,\mu_{P(n)}}
(k_{P(1)},\dots,k_{P(n)};p_{Q(1)},\dots,p_{Q(m)};q_{R(1)},\dots,q_{R(l)})
\nonumber \\
= (-1)^{|Q|+|R|} \Phi^{(nml)}_{\mu_{1},\dots,\mu_{n}}
(k_{1},\dots,k_{n};p_{1},\dots,p_{m};q_{1},\dots,q_{l}), \quad
\forall P \in {\cal P}_{n}, Q \in {\cal P}_{m}, R \in {\cal P}_{l}.
\label{sym-nml}
\end{eqnarray}

In this representation the annihilation operators have the following
expressions: 
\begin{eqnarray}
\left( A_{\nu}(r) \Phi\right)^{(nml)}_{\mu_{1},\dots,\mu_{n}}
(k_{1},\dots,k_{n};p_{1},\dots,p_{m};q_{1},\dots,q_{l}) = 
\nonumber \\ \sqrt{n+1}
\Phi^{(n+1,ml)}_{\nu,\mu_{1},\dots,\mu_{n}}
(r,k_{1},\dots,k_{n};p_{1},\dots,p_{m};q_{1},\dots,q_{l})
\end{eqnarray}
\begin{eqnarray}
\left( b(r) \Phi\right)^{(nml)}_{\mu_{1},\dots,\mu_{n}}
(k_{1},\dots,k_{n};p_{1},\dots,p_{m};q_{1},\dots,q_{l}) = 
\nonumber \\ \sqrt{m+1}
\Phi^{(n,m+1,l)}_{\mu_{1},\dots,\mu_{n}}
(k_{1},\dots,k_{n};r,p_{1},\dots,p_{m};q_{1},\dots,q_{l})
\end{eqnarray}
and
\begin{eqnarray}
\left( c(r) \Phi\right)^{(nml)}_{\mu_{1},\dots,\mu_{n}}
(k_{1},\dots,k_{n};p_{1},\dots,p_{m};q_{1},\dots,q_{l}) = 
\nonumber \\  (-1)^{m} \sqrt{l+1}
\Phi^{(nm,l+1)}_{\mu_{1},\dots,\mu_{n}}
(k_{1},\dots,k_{n};p_{1},\dots,p_{m};r,q_{1},\dots,q_{l});
\end{eqnarray}
similar expressions can be written for the creation operators.
\begin{rem}
Let us note that the sign in the last expression is introduced such that we
have ``normal" anticommutation relations i.e. $b(p)$ anticommutes with $c(q)$.
Without this sign we will have anomalous commutation relations, so the
introduction of this sign is a Klein transform.
\end{rem}

Now we can give the explicit expression of the supercharge in this
representation; starting from the definition (\ref{supercharge}) we immediately
get: 
\begin{eqnarray}
\left(Q \Phi\right)^{(nml)}_{\mu_{1},\dots,\mu_{n}}
(k_{1},\dots,k_{n};p_{1},\dots,p_{m};q_{1},\dots,q_{l}) =
\nonumber \\
(-1)^{m} \sqrt{n+1\over l} \sum_{s=1}^{l} (-1)^{s-1} q^{\nu}_{s}
\Phi^{(n+1,m,l-1)}_{\nu,\mu_{1},\dots,\mu_{n}}
(q_{s},k_{1},\dots,k_{n};p_{1},\dots,p_{m};q_{1},\dots,\hat{q_{s}},\dots,q_{l})
\nonumber \\
- \sqrt{m+1\over n} \sum_{s=1}^{n} (k_{s})_{\mu_{s}}
\Phi^{(n-1,m+1,l)}_{\mu_{1},\dots,\hat{\mu_{s}},\dots,\mu_{n}}
(k_{1},\dots,\hat{k_{s}},\dots,k_{n};k_{s},p_{1},\dots,p_{m};q_{1},\dots,q_{l})
\label{Q-explicit}
\end{eqnarray}
where, of course, we use Bourbaki convention 
$\sum_{\emptyset} \equiv 0$.

\begin{rem}
Let us note that the relations (\ref{com-dagger}) and (\ref{Q-0}) are
{\it uniquely} determining the expression of the supercharge. Indeed, one can
write any element of the Hilbert space 
${\cal H}^{gh}$
in the form
\begin{eqnarray}
\Phi = \sum_{n,m,l=0}^{\infty} {(-1)^{n}\over \sqrt{n!m!l!}}
\int_{(X^{+}_{0})^{n+m+l}} 
d\alpha^{+}_{0}(K) d\alpha^{+}_{0}(P) d\alpha^{+}_{0}(Q) 
\Phi^{(nml)}_{\mu_{1},\dots,\mu_{n}}(K,P,Q) 
\nonumber \\
A^{\dagger\mu_{1}}(k_{1})\cdots A^{\dagger\mu_{n}}(k_{n})
b^{*}(p_{1})\cdots b^{*}(p_{m})
c^{*}(q_{1})\cdots c^{*}(q_{l}) \Phi_{0}.
\end{eqnarray}
We apply the supercharge $Q$ and we commute it (using the relations
(\ref{com-dagger})) till it gives zero on the vacuum. In this way the formula
(\ref{Q-explicit}) is produced and formula (\ref{com}) becomes a consequence.
\end{rem}

Now we introduce on 
${\cal H}^{gh}$
a {\it Krein operator} according to:
\be
\left( J\Phi\right)^{(nml)}(K;P;Q) \equiv 
(-1)^{ml} (-g)^{\otimes n} \Phi^{(nlm)}(K;Q;P).
\label{Krein-gh}
\ee

In words, this operators acts on the ``photon" variables as the Krein operator
introduced in the previous subsection through the formula (\ref{Krein-em})
and, moreover,  inverts the r\^oles of the two types of ghosts (up to a sign).
The properties of this operator are summarised in the following proposition
which can be proved by elementary computations:
\begin{prop}
The following relations are verified:
\be
J^{*} = J^{-1} = J
\ee
\be
J b(p) J = c(p), \quad J c(p) J = b(p), \quad J A_{\mu}(p) J = A_{\mu}(p)
\ee
\be
J Q J = Q^{*}
\ee
and
\be
{\cal U}_{g} J = J {\cal U}_{g}, \quad \forall g \in {\cal P}.
\label{UJ}
\ee
Here $O^{*}$ is the adjoint of the operator $O$ with respect to the scalar
product 
$<\cdot,\cdot>$
on
${\cal H}^{gh}$.
\end{prop}

We can define now the sesquilinear form on 
${\cal H}^{gh}$ according to
\be
(\Psi,\Phi) \equiv <\Psi,J\Phi>;
\label{sesqui-gh}
\ee
then this form is non-degenerated.  It is convenient to denote the conjugate of
the arbitrary operator $O$ with respect to the sesquilinear form
$(\cdot,\cdot)$
by
$O^{\dagger}$ 
i.e.
\be
(O^{\dagger} \Psi,\Phi) = (\Psi,O \Phi).
\ee

Then the following formula is available:
\be
O^{\dagger} = J O^{*} J.
\ee

As a consequence, we have
\be
A_{\mu}(x)^{\dagger} = A_{\mu}(x), \quad
u(x)^{\dagger} = u(x), \quad
\tilde{u}(x)^{\dagger} = - \tilde{u}(x).
\label{conjugate}
\ee

From (\ref{UJ}) it follows that we have:
\be
({\cal U}_{g}\Psi,{\cal U}_{g}\Phi) = (\Psi,\Phi), 
\forall g \in {\cal P}^{\uparrow}, \quad
({\cal U}_{I_{t}}\Psi,{\cal U}_{I_{t}}\Phi) = \overline{(\Psi,\Phi)}.
\label{inv-sesqui}
\ee

Now, we concentrate on the description of the factor space 
$Ker(Q)/Im(Q)$. 
The following analysis, which is essential for the complete understanding of
the photon description and of the Yang-Mills generalisation seems to be missing
from the literature. We will construct a ``homotopy" for the supercharge $Q$.

Let us consider
$\epsilon: X^{+}_{0} \rightarrow \C^{4}$
a Borel bounded function such that
\be
\epsilon^{\mu}(k) k_{\mu} = - 1, \quad
\epsilon^{\mu}(k) \epsilon_{\mu}(k) = 0.
\label{eps}
\ee

(One can construct such a function in the following way. If
$k_{0} = (1,0,0,1) \in X^{+}_{0}$
then we define
$\epsilon(k_{0}) \equiv (-{1\over 2},0,0,{1\over 2})$. Now, let
$L_{k}$
be a Wigner rotation, i.e. a Borel map 
$L$
from
$X^{+}_{0}$
into the Lorentz group, such that
$L_{k}\cdot k_{0} = k$.
If we define
$\epsilon(k) \equiv L_{k}\cdot f(k_{0})$
then we have the properties \ref{eps})).

Now we give the expression for the ``homotopy" operator. We have the following
result which follows by direct computations:
\begin{prop}
Let us define the operator
\be
\tilde{Q} \equiv \int_{X^{+}_{0}} d\alpha^{+}_{0}(q) \epsilon^{\mu}(k)
\left[ A_{\mu}(k) b^{*}(k) + A^{\dagger}_{\mu}(k) c(k)\right]
\label{tildeQ}
\ee

Then the following relation is valid:
\be
Y \equiv \{Q,\tilde{Q}\} = N_{b} + N_{c} + X
\label{Y}
\ee
where 
$N_{b}$
($N_{c}$)
are particle number operators for the ghosts of type $b$ (resp. $c$) and
\be
X \equiv \int_{X^{+}_{0}} d\alpha^{+}_{0}(q) F^{\mu\nu}(k)
A^{\dagger}_{\mu}(k) A_{\nu}(k);
\label{X}
\ee
here we have introduced the notation:
\be
F^{\mu\nu}(k) \equiv k^{\mu} \epsilon^{\nu}(k) + k^{\nu} \epsilon^{\mu}(k).
\ee

Moreover the following relations are true:
\be
\tilde{Q}^{2} = 0
\ee
and
\be
[Y, Q] = 0, \quad [Y, \tilde{Q}] = 0.
\label{YQ}
\ee
\end{prop}

We call the operator 
$\tilde{Q}$
the {\it homotopy} of $Q$. If the operator $Y$ would be invertible, then we
could immediately conclude that the cohomology of the operator $Q$ is trivial.
Fortunately this is not true. However, we have:
\begin{prop}
The operator
$\left. Y\right|_{{\cal H}_{nml}}$ is invertible iff 
$m + l > 0$.
\end{prop}

{\bf Proof:}
First we need another expression for the operator $X$ defined by (\ref{X}). It
is not very difficult to prove that one has:
\be
X = A \otimes {\bf 1}
\ee 
where the operator $A$ acts only on the Bosonic variables and is given by the
expression
\be
A = d\Gamma(P);
\ee
here
$d\Gamma$
is the familiar Cook functor \cite{Co} defined by;
\be
d\Gamma(P) \psi_{1} \otimes \cdots \otimes \psi_{n} \equiv
P\psi_{1} \otimes \psi_{2} \cdots \otimes \psi_{n} + \cdots 
\psi_{1} \otimes \psi_{2} \cdots \otimes P\psi_{n} 
\ee
and the operator $P$ is in our case given by:
\be
(P \psi)_{\mu}(k) \equiv - F_{\mu\nu}(k) \psi^{\nu}(k).
\ee

In particular we immediately obtain that $P$ is a projector i.e.
$P^{2} = P$
and we have the direct sum decomposition of the one-particle Bosonic subspace
into the direct sum of
$Ran(P)$
and
$Ran(1-P)$.
Let us consider a basis in the one-particle Bosonic subspace formed by a basis
$f_{i}, \quad i \in \N$
of 
$Ran(P)$
and a basis
$g_{i}, \quad i \in \N$
of
$Ran(1-P)$.

It is clear that a basis in the $n^{\rm th}$-particle Bosonic subspace is of
the form:
$$
f_{i_{1}} \vee \cdots f_{i_{r}} \vee g_{j_{1}} \vee \cdots \vee g_{j_{s}},
\quad r, s \in \N, \quad r + s = n.
$$

Applying the operator $A$ to such a vector gives the same vector multiplied by
$s$. So, in the basis chosen above, the operator $A$ is diagonal with diagonal
elements from $\N$. It follows that the operator 
$\left. Y\right|_{{\cal H}_{nml}}$ 
can also be exhibited into a diagonal form with diagonal elements of the form
$m + l + s, \quad s \in \N$.
It is obvious now that for
$m + l > 0$
this is an invertible operator.
$\qed$

We have the following corollary:
\begin{cor}
Let us define
$
{\cal H}_{0} \equiv \oplus_{n \geq 0} {\cal H}_{n00}
$
and
$
{\cal H}_{1} \equiv \oplus_{n \geq 0, m+l > 0} {\cal H}_{nml}
$.
Then the operator $Y$ has the block-diagonal form
\begin{eqnarray}
Y = \left( \matrix{ Y_{1} & 0 \cr 0 & Y_{0} } \right) 
\end{eqnarray}
with 
$Y_{1}$
an invertible operator.
\end{cor}

Now we have the fundamental result
\begin{prop}
There exists the following vector spaces isomorphism:
\be
Ker(Q)/Im(Q) \simeq {\cal H}'/{\cal H}''
\label{factor}
\ee
where the subspaces 
${\cal H}'$
and
${\cal H}''$
have been defined in the previous subsection (see the lemmas \ref{h'1} and
\ref{h''2} respectively).
\label{cohomology}
\end{prop}

{\bf Proof:}
(i) We note that the operators $Q$ and $\tilde{Q}$ have the block-diagonal form
\begin{eqnarray}
Q = \left( \matrix{ Q_{11} & Q_{10} \cr Q_{01} & 0 } \right),
\quad
\tilde{Q} = \left( \matrix{
\tilde{Q}_{11} & \tilde{Q}_{10} \cr \tilde{Q}_{01} & 0 } \right) 
\end{eqnarray}
and from the relations (\ref{YQ}) it easily follows that we have
\be
[Y_{1}, Q_{11}] = 0
\label{Y1Q11}
\ee
\be
Y_{1} Q_{10} = Q_{10} Y_{0}, \quad Y_{0} Q_{01} = Q_{01} Y_{1}
\ee
and similar relations for the block-diagonal elements of the homotopy operator
$\tilde{Q}$. In particular we have
\be
[Y_{1}, Q_{10} \tilde{Q}_{01} ] = 0.
\label{Y1Q10}
\ee

(ii) Let now
$\Phi \in Ker(Q)$.
If we apply the relation to (\ref{Y}) the vector $\Phi$ we obtain:
\be
Y \Phi = Q \Psi
\label{Phi}
\ee
where we have defined
\be
\Psi \equiv \tilde{Q} \Phi.
\ee

If we use the block-decomposition form for the vectors $\Phi$ and $\Psi$ we
have in particular, from this relation that $$
\Psi_{0} = \tilde{Q}_{01} \Phi_{1}.
$$

If we use this relation in (\ref{Phi}) we obtain in particular that
\be
Y_{1} \Phi_{1} = Q_{11} \Psi_{1} + Q_{10} \tilde{Q}_{01} \Phi_{1}.
\ee

Because the operator
$Y_{1}$
is invertible, we have from here
\be
\Phi_{1} = Y_{1}^{-1} Q_{11} \Psi_{1} + 
Y_{1}^{-1} Q_{10} \tilde{Q}_{01} \Phi_{1}.
\ee

But from (\ref{Y1Q11}) and (\ref{Y1Q10}) we immediately obtain
$$
[Y_{1}^{-1}, Q_{11}] = 0, \quad 
[Y_{1}^{-1}, Q_{10}\tilde{Q}_{01}] = 0
$$
so the preceding relations becomes:
\be
\Phi_{1} =  Q_{11} Y_{1}^{-1} \Psi_{1} + 
Q_{10} \tilde{Q}_{01} Y_{1}^{-1} \Phi_{1}.
\ee

Now we define the vector $\psi$ by its components:
\be
\psi_{1} \equiv Y_{1}^{-1} \Psi_{1}, \quad 
\psi_{0} \equiv \tilde{Q}_{01} Y_{1}^{-1} \Phi_{1}
\ee
and we get by a simple computation
\be
\Phi - Q\psi = \left( \matrix{ 0 \cr \phi_{0} } \right) 
\ee
where
$$
\phi_{0} \equiv \Phi_{0} - Q_{01} \psi_{1}.
$$

In other words, if 
$\Phi \in Ker(Q)$
then we have
\be
\Phi = Q\psi + \tilde{\Phi}
\ee
where
\be
\tilde{\Phi}^{(nml)} = 0, \quad m + l > 0.
\ee

(iii) The condition 
$Q\Phi = 0$
amounts now to
$Q\tilde{\Phi} = 0$
or, with the explicit expression of the supercharge (\ref{Q-explicit}):
\be
q^{\nu} \tilde{\Phi}^{(n+1,0,0)}_{\nu,\mu_{1},\dots,\mu_{n}}(q,k_{1},\dots,
k_{n};\emptyset;\emptyset) = 0,
\quad \forall n \in \N
\ee
i.e. the ensemble
$\left. \{\tilde{\Phi}^{(n00)}\}\right|_{n \in \N}$
is an element from
${\cal H}'$
(see lemma \ref{h'1}).

It remains to see in what conditions such 
$\tilde{\Phi}$
is an element from 
$Im(Q)$
i.e. we have
$\tilde{\Phi} = Q \chi$.
It is clear that only the components
$\chi^{(n10)}$
should be taken non-null. Then the expression of the supercharge
(\ref{Q-explicit}) gives for any 
$n \in \N$ 
the following expressions:
\begin{eqnarray}
\left(Q \chi\right)^{(n00)}_{\mu_{1},\dots,\mu_{n}}
(k_{1},\dots,k_{n};\emptyset;\emptyset) =
- {1\over \sqrt{n}} \sum_{s=1}^{n} (k_{s})_{\mu_{s}}
\chi^{(n-1,1,0)}_{\mu_{1},\dots,\hat{\mu_{s}},\dots,\mu_{n}}
(k_{1},\dots,\hat{k_{s}},\dots,k_{n};k_{s};\emptyset)
\label{Q-chi}
\end{eqnarray}
and
\be
\left(Q \chi\right)^{(n11)}_{\mu_{1},\dots,\mu_{n}}
(k_{1},\dots,k_{n};p;q) = - \sqrt{n+1}
q^{\nu} \chi^{(n+1,1,0)}_{\nu,\mu_{1},\dots,\mu_{n}}(q,k_{1},\dots,k_{n};p;
\emptyset).
\ee

But we must have
$\left(Q \chi\right)^{(n11)} = 0$
because the vector
$Q\chi = \tilde{\Phi}$
has only the projection on 
${\cal H}_{0}$
non-null. This means that the expression (\ref{Q-chi}) is an element from 
${\cal H}_{n}''$
(see (\ref{h-secund})).
The isomorphism from the statement is now
$$
[\Phi] \leftrightarrow [\tilde{\Phi}]
$$
where in the left hand side we take classes modulo 
$Im(Q)$
and in the right hand side we take classes modulo
${\cal H}''$.
$\qed$

\begin{rem}
The homotopy formula derived above is considerably more complicated that the
one appearing in \cite{RR}).
\end{rem}

We now have a standard result (see e.g. \cite{RR}):
\begin{lemma}
The sesquilinear form 
$(\cdot,\cdot)$
induces a strictly positive defined scalar product on the factor space
$\overline{Ker(Q)/Im(Q)}$.
\label{RR}
\end{lemma}

{\bf Proof:}
One must first show that the formula
\be
([\Psi],[\Phi]) \equiv (\Psi,\Phi)
\label{positive-prod}
\ee
gives a well defined expression, i.e. the right hand side does not depend on
the representatives
$\Psi \in [\Psi]$
and
$\Phi \in [\Phi]$;
here
$[\Psi], [\Phi]$
are arbitrary equivalence classes from
$Ker(Q)/Im(Q)$.
Next, one applies the preceding proposition and can take into the right hand 
side of (\ref{positive-prod}) 
$\Psi, \Phi \in {\cal H}'$.
But in this case the sesquilinear form defined through (\ref{positive-prod})
is positive defined because it coincides with the scalar product defined in 
proving proposition \ref{photon-factor}.
$\qed$

We also have:
\begin{lemma}
The representation of ${\cal U}$ of the Poincar\'e group factors out at
$Ker(Q)/Im(Q)$.
\end{lemma}

{\bf Proof:}
One combines the results (\ref{UQ}) and (\ref{inv-sesqui}).
$\qed$

The central result follows immediately:
\begin{thm}
The isomorphism (\ref{factor}) extends to a Hilbert space isomorphism:
$$
\overline{Ker(Q)/Im(Q)} \simeq {\cal F}_{photon}.
$$
\label{photon+ghosts}
\end{thm}

\begin{rem}
It is obvious that the whole construction of the photon Fock space as a factor
space relies heavily on the property (\ref{square}) of the supercharge. This
property follows in turn from the choice of the ``wrong" statistics for the
ghosts fields i.e. Fermi-Dirac statistics. It is highly doubtful if a
supercharge with all the required properties can be constructed using Bosonic
ghosts. Nevertheless, one cannot claim that the construction above is unique.
In fact, as it appears from the literature, there are many other ``gauges" i.e.
possibilities of obtaining a theorem of the type \ref{photon+ghosts}. It is an
interesting, although not very well posed problem, to try to describe the most
general quantization of the electromagnetic field of this kind.
\end{rem}
\newpage
\subsection{Gauge-Invariant Observables\label{gau-obs}}

In this subsection we analyse in detail the same problem which was analysed at
the end of the first subsection, namely the construction of observables on the
factor space from the theorem \ref{photon+ghosts}. We hope to clarify some
point not very well treated in the literature.

First we determine by direct calculus the following relations:
\be
\{Q,u^{(\pm)}(x)\} = 0,\quad 
\{Q,\tilde{u}^{(\pm)}(x)\} =  - i \partial^{\mu} A^{(\pm)}_{\mu}(x),\quad
[Q, A^{(\pm)}_{\mu}(x)] = i \partial_{\mu} u^{(\pm)}(x);
\ee
as a consequence:
\be
\{Q,u(x)\} = 0,\quad 
\{Q,\tilde{u}(x)\} =  - i \partial^{\mu} A_{\mu}(x),
[Q, A_{\mu}(x)] = i \partial_{\mu} u(x).
\label{Q-com}
\ee

\begin{rem}
The fact that the commutation with $Q$ does not mix positive and negative parts
of the various fields follows from the clever definition of $u$ and 
$\tilde{u}$. If one would take, for instance, instead of
$\tilde{u}$
the field
$u^{*}$
such nice relations would be lost.
\end{rem}

Next, we denote by ${\cal W}$ the linear space of all Wick monomials on the 
Fock space
${\cal H}^{gh}$
i.e. containing the fields 
$A_{\mu}(x),~ u(x)$
and
$\tilde{u}(x)$.
If $M$ is such a Wick monomial, we define by
$gh_{\pm}(M)$
the degree in $\tilde{u}$ (resp. in $u$). The total degree of $M$ is obviously
\be
deg(M) \equiv gh_{+}(M) + gh_{-}(M).
\ee

The {\it ghost number} is, by definition, the expression:
\be
gh(M) \equiv gh_{+}(M) - gh_{-}(M).
\ee

Then we have the following well-known result:
\begin{lemma}
If 
$M \in {\cal W}$
let us define the operator:
\be
d_{Q} M \equiv :QM: - (-1)^{gh(M)} :MQ:
\label{BRST-op}
\ee
on monomials $M$ and extend it by linearity to the whole ${\cal W}$. Then
$d_{Q}M \in {\cal W}$
and
\be
gh(d_{Q}M) = gh(M) -1.
\ee
\end{lemma}

{\bf Proof:}
It is done by induction on 
$deg(M)$.
First, we consider the case when $M$ contains only the ghost field $u$ and
prove easily by induction on 
$gh_{-}(M)$
that $d_{Q}M = 0$. 
Indeed, if 
$deg(M) = 1$
we obviously have from the first relation (\ref{Q-com}) this equality. If we
assume that the equality is true for $M$ of degree $n$ in $u$, let $M$ be of
degree $n+1$. We can write it as follows: 
$M = :BC:$
where $B$ (resp. $C$) are Wick monomials of degree $1$ (resp. $n$) in $u$. We 
have from here:
$$
M = B^{+}C + (-1)^{n} C B^{-}
$$
and the commutator of $Q$ with this expression can be computed without any
problem, producing the desired result if one uses the induction hypothesis.

Next, we take $M$ to be a monomial in $u$ and 
$A_{\mu}$
and prove the assertion from the statement by induction on the degree in
$A_{\mu}$
in the same way. Finally, one considers the general case applying the same
tricks. 
$\qed$

The operator 
$d_{Q}: {\cal W} \rightarrow {\cal W}$
is called the {\it BRST operator}; other properties of this
object are summarized in the following elementary:
\begin{prop}
The following relations are verified:
\be
d_{Q}^{2} = 0,
\label{Q2}
\ee
\be
d_{Q} u = 0, \quad d_{Q} \tilde{u} = - i \partial^{\mu} A_{\mu}, \quad
d_{Q} A_{\mu} = i \partial_{\mu} u;
\label{BRST}
\ee
\be
d_{Q}(MN) = (d_{Q}M) N + (-1)^{gh(M)} M (d_{Q}N), \quad
\forall M, N \in {\cal W}.
\label{Leibnitz}
\ee
\end{prop}

Now we can distinguish a class of observables on the factor space from theorem
\ref{photon+ghosts}; we have in complete analogy to the construction
(\ref{O-inv}) the following result:
\begin{lemma}
If 
$O: {\cal H}^{gh} \rightarrow {\cal H}^{gh}$
verifies the condition
\be
d_{Q} O = 0
\label{dQ}
\ee
then it induces a well defined operator 
$[O]$
on the factor space
$\overline{Ker(Q)/Im(Q)} \simeq {\cal F}_{photon}$.

Moreover, in this case the following formula is true for the matrix elements of
the factorized operator 
$[O]$:
\be
([\Psi], [O] [\Phi]) = (\Psi, O \Phi). 
\label{matrix-elem}
\ee
\end{lemma}

This kind of observables on the physical space will also be called {\it gauge
invariant observables}. Next, we have in analogy to lemma \ref{gi}:
\begin{lemma}
An operator
$O: {\cal H}^{gh} \rightarrow {\cal H}^{gh}$
induces a gauge invariant observables if and only if it verifies:
\be
\left. d_{Q} O \right|_{Ker(Q)} = 0.
\ee
\label{gi-ghosts}
\end{lemma}

The criteria from lemma \ref{gi-ghosts} can be applied, as in the first
subsection, to the electromagnetic field strength (\ref{F-mu-nu}) and to the
electromagnetic potentials (\ref{em-potentials}) proving that they are
gauge-invariant observables.

Not all operators verifying the condition (\ref{dQ}) are interesting. In fact,
we have from (\ref{Q2}):
\begin{lemma}
The operators of the type
$d_{Q} O$
are inducing a null operator on the factor space; explicitly, we have:
\be
[d_{Q} O] = 0.
\ee
\label{trivial}
\end{lemma}

Moreover, we have:
\begin{thm}
Let the interaction Lagrangian be a Wick monomial
$T_{1} \in {\cal W}$
with
$gh(T_{1}) \not= 0$.
Then the chronological product are null, i.e. there is no non-trivial 
$S$-matrix.
\label{gh(M)=0}
\end{thm}
{\bf Proof:}
The generic form of 
$T_{1}$ is
$$
T_{1}(x) = :u_{1}^{\sharp}(x)\cdots u_{n}^{\sharp}(x) M(x):
$$
where $M$ is a Wick monomial in 
$A_{\mu}(x)$.
One must determine the corresponding chronological products
$T_{n},~n \geq 2$. 
It is not very hard to prove, by induction, that we also have:
\be
gh(T_{n}) \not= 0.
\label{gh(T)}
\ee

Indeed, to compute
$T_{2}$
one splits causally the commutator
$D_{2}(x,y) = [T_{1}(x),T_{1}(y)]$.
But one can compute this expression using Wick theorem. From 
(\ref{CAR-ghosts}) one can see that the only non-null pairing are between a
ghost of type $u$ and a ghost of type 
$\tilde{u}$.
So, we have
$gh(D_{2}) \not= 0$.
It is clear that this property is preserved by the process of distribution
splitting, so we also have
$gh(T_{2}) \not= 0$.
The argument goes on now to an arbitrary order producing chronological product
verifying (\ref{gh(T)}).  

We compute now the matrix elements of 
$T_{n}$ using
(\ref{matrix-elem}); according to proposition \ref{cohomology} we can take into
the right hand side,
$\Psi, \Phi$ of the form
$\Psi = \Psi' \otimes \Psi_{0}^{gh}$ and
$\Phi = \Phi' \otimes \Psi_{0}^{gh}$
where
$\Psi_{0}^{gh}$
is the vacuum state in the ghost Hilbert space and
$\Psi', \Phi'$
are arbitrary vectors from 
${\cal H}$.
It follows that we have:
$$
([\Psi], [T_{n}(x_{1},\dots,x_{n})][\Phi])  = 
\sum_{\alpha} (\Phi^{gh}_{0}, G^{\alpha}_{n} \Phi^{gh}_{0}) 
(\Psi',M^{\alpha}_{n}\Phi')
$$
where 
$G^{\alpha}_{n}$
are Wick monomials in the ghost fields and
$M^{\alpha}_{n}$
are Wick monomials in the electromagnetic potentials. According to
(\ref{gh(T)}) we have
$gh(G^{\alpha}_{n}) \not= 0$.
But, in this case one can compute 
$(\Phi^{gh}_{0}, G^{\alpha}_{n} \Phi^{gh}_{0})$
using Wick theorem (see the end of subsection \ref{ff}) and (\ref{CAR-ghosts});
because only the pairing of $u$ and
$\tilde{u}$
can give non-null contributions, it follows that, in the end we get
$([\Psi], [T_{n}(x_{1},\dots,x_{n})][\Phi])  = 0$.
$\qed$

As in the end of subsection \ref{quant} one can see that interaction
Lagrangians of the type (\ref{interaction}) do not factorise to the ``physical"
space
$Ker(Q)/Im(Q)$.
Finally, we want to analyse possible interactions between the electromagnetic
field and ``matter" in this new representation of the electromagnetic field.
Presumably, we will obtain the same result as in the end of subsection 
\ref{quant}. Indeed, if 
${\cal H}_{matter}$
is the corresponding Hilbert space of the matter fields, it is elementary to 
see that we can realise the total Hilbert space
${\cal H}_{total} \equiv {\cal F}_{photon} \otimes {\cal H}_{matter}$
as the factor space 
$Ker(Q)/Im(Q)$
where the supercharge $Q$ is defined on
$\tilde{\cal H}_{gh} \equiv {\cal H}_{gh} \otimes {\cal H}_{matter}$
by the obvious substitution
$Q \rightarrow Q \otimes {\bf 1}$.

Now we have:
\begin{thm}
Let us define on
$\tilde{\cal H}_{gh}$
the interaction Lagrangian of the form (\ref{inter}) where the current
$j^{\mu}(x)$
is a Wick monomial in the matter and ghost fields. The this expression
factorises, in the adiabatic limit, to the physical space
${\cal H}_{total}$
and gives a non-null $S$-matrix if and only if it does not depend on the ghost
fields and it is conserved in the sense (\ref{conservation-1}).
\end{thm}

{\bf Proof:}
We apply the criterion from lemma \ref{gi-ghosts} to the interaction Lagrangian
(\ref{inter}) and easily obtain the conservation law (\ref{conservation-1})
{\it and} the condition
$
d_{Q} j^{\mu}(x) = 0.
$
Using the relations (\ref{BRST}) it is not hard to prove that the current
$j^{\mu}(x)$
should not contain ghosts of the type 
$\tilde{u}$.
But in this case one can apply proposition \ref{gh(M)=0} and obtain that, in
fact, the current does not depend on $u$ either.
$\qed$
\newpage

\section{Pure Yang-Mills Fields\label{ym}}

\subsection{The General Framework}

In this section, we derive the following result: the only possible coupling
between 
$r > 1$
electromagnetic-type fields of power not greater than $4$ is through an
Yang-Mills Lagrangian. This type of result has recently appeared in \cite{AS}.
The main differences in our approach is the systematic utilisation of the
criterion from lemma \ref{gi-ghosts} ; we also hope to streamline the
arguments.

First, we have to define in an unambiguous way what we mean by Yang-Mills
fields. We make in subsection \ref{gh} the following modifications. 

$\bullet$
Instead of 
${\sf H}$ 
(see subsection \ref{ff}) we consider Hilbert space
${\sf H}^{r} \equiv L^{2}(X^{+}_{0},(\C^{4})^{r},d\alpha_{m}^{+})$
with the scalar product
\be
<\phi,\psi> \equiv \sum_{a=1}^{r} \int_{X^{+}_{0}} d\alpha_{0}^{+} 
<\phi_{a}(p),\psi_{a}(p)>_{\C^{4}};
\ee
The subspaces
$
{\sf H}_{YM}'\equiv 
\{ \phi \in {\sf H}^{r} \vert \quad p^{\mu} \phi_{a\mu}(p) = 0\}
$
and
$
{\sf H}_{YM}'' \equiv 
\{ \phi \in {\sf H}_{YM}' \vert \quad \Vert\phi\Vert = 0\}
$
are introduced by complete analogy and we call the factor space
$
{\sf H}_{YM}^{r} \equiv \overline{{\sf H}_{YM}'/{\sf H}_{YM}''}
$
the Hilbert space of the {\it Yang-Mills} particles (or {\it gluons}). The
description of the corresponding Fock space can be done as in subsection
\ref{quant}. 

$\bullet$
The construction of the ghosts fields starts from 
${\sf H}^{gh,r} \equiv L^{2}(X^{+}_{m},(\C^{2})^{r},d\alpha_{m}^{+})$
with the scalar product
\be
<\phi,\psi> \equiv \sum_{a=1}^{r} \int_{X^{+}_{0}} d\alpha_{0}^{+} 
<\phi_{a}(p),\psi_{a}(p)>_{\C^{2}};
\ee
the corresponding Fock space is called the space of {\it ghosts particles}. 

$\bullet$
The construction of the Yang-Mills interaction will be done in the Hilbert
space
$
{\cal H}^{gh,r}_{YM} \equiv {\cal H}_{YM}^{r} \otimes 
{\cal F}^{-}({\sf H}^{gh,r}).
$
One can decompose this Fock space into subspaces with fixed number of gluons
and ghosts.

$\bullet$
One can define now the Yang-Mills and the ghosts fields
$A_{a\mu}^{\sharp},~ u_{a},~\tilde{u}_{a}, \quad a = 1,\dots,r$
generalising in an obvious way the formul\ae~ (\ref{em}) + (\ref{em-pm}):
\be
A_{a\mu}(x) \equiv {1\over (2\pi)^{3/2}} \int_{X^{+}_{0}} d\alpha^{+}_{0}(p) 
\left[ e^{-ip\cdot x} A_{a\mu}(p) + e^{ip\cdot x} A^{\dagger}_{a\mu}(p)\right]
\label{YM}
\ee
and respectively (\ref{u}) + (\ref{u-tilde}):
\be
u_{a}(x) \equiv {1\over (2\pi)^{3/2}} \int_{X^{+}_{0}} d\alpha^{+}_{0}(q)
\left[ e^{-i q\cdot x} b_{a}(q) + e^{i q\cdot x} c_{a}^{\dagger}(q) \right]
\label{u-YM}
\ee
and
\be
\tilde{u}_{a}(x) \equiv {1\over (2\pi)^{3/2}} 
\int_{X^{+}_{0}} d\alpha^{+}_{0}(q)
\left[ - e^{-i q\cdot x} c_{a}(q) + e^{i q\cdot x} b_{a}^{\dagger}(q) \right].
\label{u-tilde-YM}
\ee

$\bullet$
The only significant modification appears in the canonical (anti)commutation
relations (\ref{CCR-em-pm}) + (\ref{CCR-em}) and (\ref{CAR-ghosts}), 
namely we have:
\be
\left[A^{(\mp)}_{a\mu}(x),A^{(\mp)}_{b\nu}(y)\right] = 
- \delta_{ab} g_{\mu\nu} D^{(+)}_{0}(x-y) \times {\bf 1},\quad
\left[A_{a\mu}(x),A_{b\nu}(y)\right] = - 
\delta_{ab} g_{\mu\nu} D_{0}(x-y) \times {\bf 1}
\label{CCR-YM}
\ee
and
\be
\{ u_{a}^{(\epsilon)}(x),\tilde{u}_{b}^{(-\epsilon)}(y)\} = 
\delta_{ab} D^{(-\epsilon)}_{0}(x-y), \quad
\{u_{a}(x),\tilde{u}_{b}(y)\} = \delta_{ab} D_{0}(x-y).
\label{CAR-ghosts-YM}
\ee

$\bullet$
The supercharge is given by (see (\ref{supercharge})):
\be
Q \equiv \sum_{a=1}^{r} \int_{X^{+}_{0}} d\alpha^{+}_{0}(q) k^{\mu}
\left[ A_{a\mu}(k) c_{a}^{\dagger}(k) + A^{\dagger}_{a\mu}(k) b_{a}(k)\right]
\label{supercharge-YM}
\ee
and verifies all the expected properties. 

$\bullet$
The Krein operator has an expression similar to (\ref{Krein-gh}) and can be 
used to construct a sesquilinear form like in (\ref{sesqui-gh}). Then relations
of the type (\ref{conjugate}) are still true;
\be
A_{a\mu}(x)^{\dagger} = A_{a\mu}(x), \quad
u_{a}(x)^{\dagger} = u_{a}(x), \quad
\tilde{u}_{a}(x)^{\dagger} = - \tilde{u}_{a}(x).
\label{conjugate-YM}
\ee

$\bullet$
As a consequence, proposition \ref{cohomology}, and the main theorem
\ref{photon+ghosts} stay true.

$\bullet$
The ghost degree is defined in an obvious way and the expression of the BRST
operator (\ref{BRST-op}) is the same in this more general framework and the
corresponding properties are easy to obtain. In particular we have (see
(\ref{BRST})):
\be
d_{Q} u_{a} = 0, \quad d_{Q} \tilde{u}_{a} = - i \partial^{\mu} A_{\mu}, \quad
d_{Q} A_{a\mu} = i \partial_{\mu} u_{a}.
\label{BRST-YM}
\ee

$\bullet$
Finally, the characterisation of gauge-invariant observables is done in the
same way as in subsection \ref{gau-obs}. In particular we have the theorem
\ref{gh(M)=0}.

As we see, the Yang-Mills fields are nothing but a $r$-th component
electromagnetic-type {\it free} field. This assumption has its limitation,
but it is one of the main ingredients of the theory, although not always
admitted explicitly in the literature.

We will try to construct a $S$-matrix on the auxiliary Fock space
${\cal H}^{gh,r}_{YM}$
having all the properties from the section \ref{pt}. In particular we should be
able to construct the first-order term, i.e. the interaction Lagrangian
$T_{1}(x)$
verifying the properties (\ref{inv1}), (\ref{causality1}), (\ref{unitarity1})
and (\ref{adiabatic1}). We emphasise that in the condition (\ref{unitarity1}) 
we will consider the adjoint with respect to the sesquilinear form
$(\cdot,\cdot)$
defined with the help of the Krein operator. Moreover, we will impose that the
chronological products constructed in this way should have a well defined
adiabatic limit. According to lemma \ref{gi-ghosts} this condition writes as:
\be
lim_{\epsilon \searrow 0} \left. d_{Q} T_{n}(x_{1},\dots,x_{n})\right|_{Ker(Q)}
= 0, \quad \forall n \geq 1.
\label{gi-Tn}
\ee

If these two requirements are fulfilled, then we get an unitary $S$-matrix on
the factor space $Ker(Q)/Im(Q)$; indeed, if the chronological products are
factorizing to this quotient, the the same must be true for the
antichronological products, because they are expressed by (\ref{antichrono}) in
terms of the chronological ones. In this way, the whole unitarity argument of
Epstein and Glaser \cite{EG1} works in our case also: we have from
(\ref{unitarity1}), by induction, that (\ref{unitarity}) can be fulfilled and by
factorization we get an unitary $S$-matrix on the quotient (``physical") space.
This is due to the fact that the sesquilinear form induces on this quotient
space a true scalar product according to lemma \ref{RR}.
\newpage
\subsection{The Derivation of the Yang-Mills Lagrangian}

In this subsection we prove two theorems which characterise completely the
Yang-Mills interaction of gluons. We assume the summation convention of the
dummy indices 
$a,b,\dots.$

\begin{thm}
Let us consider the operator
\be
T_{1}(g) = \int_{\R^{4}} dx~ g(x) T_{1}(x)
\ee
defined on 
${\cal H}^{gh,r}_{YM}$
with
$T_{1}$
a Lorentz-invariant Wick polynomial in 
$A_{\mu},~u$ 
and 
$\tilde{u}$
verifying also
$\omega(T_{1}) \leq 4$.
Then 
$T_{1}(g)$
can induce an well defined non-trivial $S$-matrix, in the adiabatic limit, if
and only if it has the following form:
\begin{eqnarray}
T_{1}(g) = \int_{\R^{4}} dx~ g(x) 
\{ f_{abc} \left[ :A_{a\mu}(x)A_{b\nu}(x) \partial^{\nu} A_{a}^{\mu}(x): -
:A_{a}^{\mu}(x) u_{b}(x) \partial_{\mu} \tilde{u}_{c}(x): \right] +
\nonumber \\
h_{ab} \left[ : A_{a\mu}(x)A_{b}^{\mu}(x): -
2 :\tilde{u}_{a}(x) u_{b}(x):\right] \}.
\label{YM-1}
\end{eqnarray}
Here 
$f_{abc}$
($h_{ab}$)
are completely antisymmetric (symmetric) constants.
\label{T1}
\end{thm}

{\bf Proof:}
(i) If we take into account Lorentz invariance, the power counting condition
from the statement and the restriction of non-triviality of theorem
\ref{gh(M)=0} we end up with the following linear independent possibilities:

$\bullet$ of degree 2:
\be
T^{(1)'} = h^{(1)}_{ab} :A_{a\mu}(x)~ A_{b}^{\mu}(x):, \quad
T^{(2)'} = h^{(2)}_{ab} :\tilde{u}_{a}(x)~ u_{b}(x):
\ee

$\bullet$ of degree 3: none

$\bullet$ of degree 4:
\begin{eqnarray}
T^{(1)} = f^{(1)}_{abc} :A_{a\mu}(x)~ A_{b\nu}(x)~ 
\partial^{\nu} A_{c}^{\mu}(x): \quad 
T^{(2)} = f^{(2)}_{abc} :A_{a}^{\mu}(x)~ u_{b}(x) 
\partial_{\mu} \tilde{u}_{c}(x): \quad 
\nonumber \\
T^{(3)} = f^{(3)}_{abc} :A_{a}^{\mu}(x)~ \partial_{\mu} u_{b}(x)~ 
\tilde{u}_{c}(x): \quad
T^{(4)} = f^{(4)}_{abc} :\partial_{\mu} A_{a}^{\mu}(x)~ u_{b}(x) 
\tilde{u}_{c}(x): \quad 
\nonumber \\
T^{(5)} = f^{(5)}_{abc} :A_{a\mu}(x)~ A_{b}^{\mu}(x)~ 
\partial_{\nu} A_{c}^{\nu}(x): \quad 
T^{(6)} = g^{(1)}_{abcd} :A_{a\mu}(x)~ A_{b}^{\mu}(x)~ 
A_{c\nu}(x)~ A_{d}^{\nu}(x): 
\nonumber \\
T^{(7)} = g^{(2)}_{abcd} :A_{a\mu}(x)~ A_{b}^{\mu}(x)~ 
u_{c}(x)~ \tilde{u}_{d}(x): \quad
T^{(8)} = g^{(3)}_{abcd} :u_{a}(x)~ u_{b}(x)~ 
\tilde{u}_{c}(x)~ \tilde{u}_{d}(x): 
\nonumber \\
T^{(9)} = g^{(4)}_{abcd} \varepsilon_{\mu\nu\rho\sigma}
:A_{a}^{\mu}(x)~ A_{b}^{\nu}(x)~ A_{c}^{\rho}(x)~ A_{d}^{\sigma}(x): \quad
T^{(10)} = g^{(1)}_{ab} :\partial_{\mu} A_{a\nu}(x)~ 
\partial^{\mu} A_{b}^{\nu}(x):  
\nonumber \\
T^{(11)} = g^{(2)}_{ab} :\partial_{\mu} A_{a}^{\mu}(x)~ 
\partial_{\nu} A_{b}^{\nu}(x): \quad 
T^{(12)} = g^{(3)}_{ab} :\partial_{\mu} A_{a\nu}(x)~ 
\partial^{\nu} A_{b}^{\mu}(x): 
\nonumber \\
T^{(13)} = g^{(4)}_{ab} :A_{a}^{\mu}(x)~ 
\partial_{\mu} \partial_{\nu} A_{b}^{\nu}(x): \quad 
T^{(14)} = g^{(5)}_{ab} \varepsilon_{\mu\nu\rho\sigma}
:F_{a}^{\mu\nu}(x)~ F_{b}^{\rho\sigma}(x): 
\nonumber \\
T^{(15)} = g^{(6)}_{ab} :\partial_{\mu} u_{a}(x)~ 
\partial^{\mu} \tilde{u}_{b}(x):
\label{list}
\end{eqnarray}

Without losing generality we can impose the following symmetry restrictions on
the constants from the preceding list:
\begin{eqnarray}
h^{(1)}_{ab} = h^{(1)}_{ba} \quad
g^{(1)}_{abcd} = g^{(1)}_{bacd} = g^{(1)}_{abdc} = g^{(1)}_{cdab} \quad
g^{(2)}_{abcd} = g^{(2)}_{bacd} 
\nonumber \\
g^{(3)}_{abcd} = - g^{(3)}_{bacd} = - g^{(3)}_{abdc}  \quad
g^{(i)}_{ab} = g^{(2)}_{ba} \quad i = 1, 2, 3, 5
\label{s1}
\end{eqnarray}
and one can suppose that
$g^{(4)}_{abcd}$
are completely antisymmetric in all indices. 

(ii) By integration over $x$ some of the linear independence is lost in the
adiabatic limit. Namely:

$\bullet$
One can eliminate
$T^{(3)}$
by redefining the constants
$f^{(2)}_{abc}$
and
$f^{(4)}_{abc}$;

$\bullet$
One can eliminate
$T^{(5)}$
by redefining the constants
$f^{(1)}_{abc}$;

$\bullet$
One can eliminate
$T^{(12)}$
and
$T^{(13)}$
by redefining the constants
$g^{(2)}_{ab}$;

$\bullet$
One can eliminate
$T^{(10)}$
and
$T^{(15)}$
using the equation of motion (\ref{eqA}) and (\ref{equ});

$\bullet$
Finally
$T^{(14)}$
is null in the adiabatic limit.

(iii)
Some of the remaining expressions are of the form
$d_{Q} O$
so they do not count, according to lemma \ref{trivial}. Namely,
$$
T^{(11)} = i d_{Q} \left(: 
g^{(2)}_{ab} : \partial_{\mu} A^{\mu}_{a} \tilde{u}_{b}: \right)
$$
and if we define
$$
O \equiv g_{abc} :\tilde{u}_{a} u_{b} \tilde{u}_{c}:
$$
with the constants
$g_{abc}$
antisymmetric in the indices $a$ and $c$, then we have
$$
d_{Q} O = 2i g_{abc} : \partial_{\mu} A^{\mu}_{a} u_{b} \tilde{u}_{c}:
$$
so, it follows that we can choose 
\be
f^{(4)}_{abc} = f^{(4)}_{cba}.
\label{s2}
\ee

(iv) As a conclusion, we can keep in
$T_{1}$
only the expressions
$T^{(1)'},~ T^{(2)'},~ T^{(1)},~ T^{(2)},~ T^{(4)}$
and
$T^{(6)}-T^{(9)}$
with the symmetry properties (\ref{s1}) and (\ref{s2}).

We compute now the expression
$d_{Q} T_{1}$;
we find:
\begin{eqnarray}
d_{Q} T_{1} = i \partial_{\mu} [
2 h^{(1)}_{ab} :A_{a}^{\mu} u_{b}: + 
( f^{(1)}_{abc} - f^{(1)}_{cba}) :u_{a} A_{b\nu} \partial^{\nu} A^{\mu}_{c}:
+ f^{(1)}_{bac} :u_{a} A_{b\nu} \partial^{\mu} A^{\nu}_{c}: 
\nonumber \\
+ f^{(1)}_{bca} :\partial_{\nu}u_{a} A_{b}^{\nu} A^{\mu}_{c}: -
f^{(1)}_{cba} :u_{a} \partial^{\nu}A_{b\nu} A_{c\nu}: ] 
\nonumber \\
- i (2 h^{(1)}_{ab} + h^{(2)}_{ab} ) :\partial_{\mu}A_{a}^{\mu} u_{b}: 
+i ( f^{(1)}_{cba} - f^{(1)}_{abc}) 
:u_{a} \partial_{\mu}A_{b\nu} \partial^{\nu}A^{\mu}_{c}: 
\nonumber \\
+ i ( f^{(1)}_{abc} - f^{(1)}_{bac} - f^{(1)}_{cba} + f^{(2)}_{cba})
:\partial_{\mu}\partial_{\nu}A_{a}^{\mu} A_{b}^{\nu} u_{c}: +
i (f^{(1)}_{abc} + f^{(4)}_{acb})
:\partial_{\mu}A_{a}^{\mu} \partial_{\nu}A_{b}^{\nu} u_{c}: 
\nonumber \\
- i f^{(1)}_{acb}:\partial_{\nu}A_{a\mu} \partial^{\nu}A_{b}^{\mu} u_{c}: +
i f^{(2)}_{abc}:\partial^{\mu}u_{a} u_{b} \partial_{\mu}\tilde{u}_{c}: 
+ 4i g^{(1)}_{abcd} :\partial_{\mu}u_{a} A_{b}^{\mu} A_{c\nu} A_{d}^{\nu}: 
\nonumber \\
+ i g^{(2)}_{abcd} ( 2 :\partial_{\mu}u_{a} A_{b}^{\mu} u_{c} \tilde{u}_{d}: +
:A_{a\mu} A_{b}^{\mu} u_{c} \partial_{\rho} A^{\rho}_{d}: ) 
- 2 i g^{(3)}_{abcd} :u_{a} u_{b} \partial_{\mu}A_{c}^{\mu} \tilde{u}_{d}: 
\nonumber \\
-4 i g^{(4)}_{abcd} \varepsilon_{\mu\nu\rho\sigma}
:\partial^{\mu}u_{a} A_{b}^{\nu} A_{c}^{\rho} A_{d}^{\sigma}: 
\end{eqnarray}

Using lemma \ref{gi-ghosts} we impose the condition from the statement:
\be
lim_{\epsilon \searrow 0} d_{Q} 
\left. \int_{\R^{4}} dx~ g_{\epsilon}(x) T_{1}(x)\right|_{Ker(Q)} = 0.
\ee

The divergence gives no contribution and the other terms can be computed on
vectors from
${\cal H}'$
according to the proposition \ref{cohomology}. In this way we see that we get
independent conditions from each term i.e. we have:
\be
g^{(i)}_{abcd} = 0, \quad i = 1,2,3,4,
\ee
\be
2 h^{(1)}_{ab} + h^{(2)}_{ab} = 0
\ee
and
\begin{eqnarray}
f^{(1)}_{cba} - f^{(1)}_{abc} = - (b \leftrightarrow c)
\nonumber \\
f^{(1)}_{abc} - f^{(1)}_{bac} - f^{(1)}_{cba} + f^{(2)}_{cba} = 0
\nonumber \\
f^{(1)}_{abc} + f^{(4)}_{acb} = - (a \leftrightarrow b)
\nonumber \\
f^{(1)}_{acb} = - (a \leftrightarrow b).
\end{eqnarray}

We exploit completely the last system of equations. From the last relation of
the system we see that the expression
$f^{(1)}_{abc}$
is antisymmetric in $a$ and $c$. In this case, the first relation of the 
system gives us the antisymmetry in $b$ and $c$; so the expressions
$f^{(1)}_{abc}$
are completely antisymmetric in {\it all} indices. Then the second relation of
the system leads to
$$
- f^{(1)}_{cba} + f^{(2)}_{cba} = 0.
$$

At last, we use (\ref{s2}) in the third relation of the system and get that
$$
f^{(4)}_{abc} = 0.
$$

The expression from the statement have been obtained.

(vi) It remains to prove that the expression from the statement cannot be of
the type
$d_{Q} O$
and this can be done without problems.
$\qed$

\begin{rem}
The second contribution in the generic expression for 
$T_{1}$
just obtained could be left out according to the argument presented at the end
of subsection \ref{ff}. We will prefer to keep this term in the following and
show that it must be null using only arguments of gauge invariance.
\end{rem}

\begin{cor}
In the condition of the preceding theorem, one has:
\be
d_{Q} T_{1} = i\partial_{\mu} K_{1}^{\mu}
\label{k1}
\ee
where:
\be
K_{1}^{\mu} = 2 h_{ab} :A_{a}^{\mu} u_{b}: + 
f_{abc} \left( :u_{a} A_{b\nu} F^{\nu\mu}_{c}: -
{1\over 2} :u_{a} u_{b} \partial^{\mu} \tilde{u}_{c}: \right).
\ee
\end{cor}

Moreover, we have:
\begin{prop}
The expression
$T_{1}$
from the preceding theorem verifies the unitarity condition
$$
T_{1}(x)^{\dagger} = T_{1}(x)
$$
if and only if the constants 
$f_{abc}$
and
$h_{ab}$
have real values.
\end{prop}

The proof is very simple and relies on the relations (\ref{conjugate-YM}). 

Finally we have
\begin{prop}
The expression 
$T_{1}$
determined in the preceding theorem verifies the causality condition:
$$
[T_{1}(x), T_{1}(y)] = 0,\quad \forall x, y \in \R^{4} \quad {\rm s.t.} \quad
(x-y)^{2} < 0. 
$$
\end{prop}

{\bf Proof:} 
One must determine the commutator appearing in the lefthand side. After a
tedious computation one gets:
\begin{eqnarray}
D_{2}(x,y) = 
- f_{cab} f_{cde} D_{0}(x-y) [
:A_{a\nu}(x) F_{b}^{\nu\mu}(x) A_{d}^{\rho}(y) F_{e\rho\mu}(y): 
\nonumber \\
+ :u_{a}(x) \partial_{\mu}\tilde{u}_{b}(x) 
u_{d}(y) \partial^{\mu}\tilde{u}_{e}(y): 
- A_{a\nu}(x) F_{b}^{\nu\mu}(x) u_{d}(y) \partial_{\mu}\tilde{u}_{e}(y): 
\nonumber \\
- :u_{d}(x) \partial_{\mu}\tilde{u}_{e}(x) A_{a\nu}(y) F_{b}^{\nu\mu}(y): ]
+ {1\over 2} f_{abc} f_{dbc} D_{0}^{2}(x-y) :F_{a}^{\nu\mu}(x) F_{d\nu\mu}(y):
\nonumber \\
+ f_{cab} f_{cde} {\partial \over \partial x_{\rho}} D_{0}(x-y) [
:A_{a\nu}(x) F_{b}^{\nu\mu}(x) A_{d\mu}(y) A_{e\rho}(y): 
- :A_{d\mu}(x) A_{e\rho}(x) A_{a\nu}(y) F_{b}^{\nu\mu}(y): 
\nonumber \\
+ :A_{a\mu}(x) A_{b\rho}(x) u_{d}(y) \partial^{\mu}\tilde{u}_{e}(y): 
- :u_{d}(x) \partial^{\mu}\tilde{u}_{e}(x) A_{a\mu}(y) A_{b\rho}(y): 
\nonumber \\
+ :A_{a}^{\mu}(x) \partial_{\mu}\tilde{u}_{b}(x) A_{d\rho}(y) u_{e}(y): 
- :A_{d\rho}(x) u_{e}(x) A_{a}^{\mu}(y) \partial_{\mu}\tilde{u}_{b}(y): ]
\nonumber \\
- f_{cab} f_{cde} {\partial^{2} \over \partial x_{\nu} \partial x_{\rho}}
D_{0}(x-y) :A_{a\mu}(x) A_{b\nu}(x) A_{d}^{\mu}(y) A_{e\rho}(y): 
\nonumber \\
-2 f_{abc} f_{dbc} D_{0,\mu\nu}(x-y) :A_{a}^{\mu}(x) A_{d}^{\nu}(y):
+ {1\over 2} f_{abc} f_{dbc} {\partial \over \partial x_{\rho}} D_{0}^{2}(x-y) 
[:A_{a\mu}(x) F_{d}^{\mu\rho}(y): 
\nonumber \\
- :F_{d}^{\mu\rho}(x) A_{a\mu}(y): 
- :u_{a}(x) \partial_{\rho}\tilde{u}_{d}(y): 
- :\partial_{\rho}\tilde{u}_{d}(x) u_{a}(y):]
- {1\over 3} f_{abc} f_{abc} \square D_{0}^{3}(x-y) {\bf 1} 
\nonumber \\
- 4 D_{0}(x-y) h_{ab} h_{cb} [:A_{a\mu}(x) A_{c}^{\mu}(y): 
+ :u_{a}(x) \tilde{u}_{c}(y): - :\tilde{u}_{a}(x) u_{c}(y):]
\nonumber \\
+ 4h_{ab} h_{ab} D_{0}^{2}(x-y) {\bf 1}
\nonumber \\
-2 f_{cab} h_{cd} D_{0}(x-y) [:A_{a\nu}(x) F_{b}^{\nu\mu}(x) A_{d\mu}(y): 
- :u_{a}(x) \partial_{\mu}\tilde{u}_{b}(x) A_{d}^{\mu}(y): 
\nonumber \\
- :A_{a}^{\mu}(x) \partial_{\mu}\tilde{u}_{b}(x) u_{d}(y):
+ :A_{d}^{\mu}(x) u_{a}(y) \partial_{\mu}\tilde{u}_{b}(y): 
\nonumber \\
- :A_{a}^{\mu}(x) \partial_{\mu}\tilde{u}_{b}(y) u_{d}(y): 
+ :u_{d}(x) A_{a}^{\mu}(y) \partial_{\mu}\tilde{u}_{b}(y):]
\nonumber\\
-2 f_{cab} h_{cd} {\partial \over \partial x_{\nu}} D_{0}(x-y) [
:A_{a\mu}(x) A_{b\nu}(x) A_{d}^{\mu}(y):
+ :A_{d}^{\mu}(x) A_{a\mu}(y) A_{b\nu}(y):
\nonumber \\
+ :A_{a\nu}(x) u_{b}(x) \tilde{u}_{d}(y):
+ :\tilde{u}_{d}(x) A_{a\nu}(y) u_{b}(y): ]
\label{D2}
\end{eqnarray}

where
\be
D_{0}^{2}(x-y) \equiv D^{(+)}_{0}(x-y)^{2} - D^{(-)}_{0}(x-y)^{2}, \quad 
D_{0}^{3}(x-y) \equiv D^{(+)}_{0}(x-y)^{3} + D^{(-)}_{0}(x-y)^{3}
\ee
are well defined distributions with causal support and
\be
D_{0,\mu\nu}(x-y) \equiv {1\over 6} 
( \partial_{\mu} \partial_{\nu} - \square g_{\mu\nu} ) D_{0}^{2}(x-y).
\ee

The preceding expression show that we have causality in the first order.
$\qed$

We note that the following identity is valid:
\be
\partial^{\mu} D_{0,\mu\nu}(x) = 0.
\label{D-mu-nu}
\ee

We go now to the second order of the perturbation theory following closely
\cite{DHKS1}. We split causally the commutator (\ref{D2}) according to the
prescription given (\ref{split}) and (\ref{t2}) and include the most general 
finite arbitrariness of the decomposition. We can prove that a causal 
splitting which preserves Lorentz covariance of the distributions
$D_{0}(x),~D_{0}^{2}(x),~D_{0}^{3}(x)~,D_{0,\mu\nu}$
exists. Moreover, the splitting can be chosen such that it will preserve the
property (\ref{D-mu-nu}). The result is contained in:

\begin{prop}
The generic for of the distribution 
$T_{2}$
is
\begin{eqnarray}
T_{2}(x,y) = 
- f_{cab} f_{cde} D_{0,F}(x-y) [
:A_{a\nu}(x) F_{b}^{\nu\mu}(x) A_{d}^{\rho}(y) F_{e\rho\mu}(y): 
\nonumber \\
+ :u_{a}(x) \partial_{\mu}\tilde{u}_{b}(x) 
u_{d}(y) \partial^{\mu}\tilde{u}_{e}(y): 
- A_{a\nu}(x) F_{b}^{\nu\mu}(x) u_{d}(y) \partial_{\mu}\tilde{u}_{e}(y): 
\nonumber \\
- :u_{d}(x) \partial_{\mu}\tilde{u}_{e}(x) A_{a\nu}(y) F_{b}^{\nu\mu}(y): ]
+ {1\over 2} f_{abc} f_{dbc} D_{0,F}^{2}(x-y) 
:F_{a}^{\nu\mu}(x) F_{d\nu\mu}(y):
\nonumber \\
+ f_{cab} f_{cde} {\partial \over \partial x_{\rho}} D_{0,F}(x-y) [
:A_{a\nu}(x) F_{b}^{\nu\mu}(x) A_{d\mu}(y) A_{e\rho}(y): 
- :A_{d\mu}(x) A_{e\rho}(x) A_{a\nu}(y) F_{b}^{\nu\mu}(y): 
\nonumber \\
+ :A_{a\mu}(x) A_{b\rho}(x) u_{d}(y) \partial^{\mu}\tilde{u}_{e}(y): 
- :u_{d}(x) \partial^{\mu}\tilde{u}_{e}(x) A_{a\mu}(y) A_{b\rho}(y): 
\nonumber \\
+ :A_{a}^{\mu}(x) \partial_{\mu}\tilde{u}_{b}(x) A_{d\rho}(y) u_{e}(y): 
- :A_{d\rho}(x) u_{e}(x) A_{a}^{\mu}(y) \partial_{\mu}\tilde{u}_{b}(y): ]
\nonumber \\
- f_{cab} f_{cde} [{\partial^{2} \over \partial x_{\nu} \partial x_{\rho}}
D_{0,F}(x-y) + c_{1}g^{\nu\rho} \delta (x-y)] 
:A_{a\mu}(x) A_{b\nu}(x) A_{d}^{\mu}(y) A_{e\rho}(y): 
\nonumber \\
-2 f_{abc} f_{dbc} [D_{0,F,\mu\nu}(x-y) + c_{2} g_{\mu\nu} \delta (x-y)]
:A_{a}^{\mu}(x) A_{d}^{\nu}(y):
\nonumber \\
+ {1\over 2} f_{abc} f_{dbc} {\partial \over \partial x_{\rho}} 
D_{0,F}^{2}(x-y) [:A_{a\mu}(x) F_{d}^{\mu\rho}(y): 
\nonumber \\
- :F_{d}^{\mu\rho}(x) A_{a\mu}(y): 
- :u_{a}(x) \partial_{\rho}\tilde{u}_{d}(y): 
- :\partial_{\rho}\tilde{u}_{d}(x) u_{a}(y):]
- {1\over 3} f_{abc} f_{abc} \square D_{0,F}^{3}(x-y) {\bf 1} 
\nonumber \\
- 4 h_{ab} h_{cb} [D_{0,F}(x-y) + c_{3} \delta (x-y)] 
[:A_{a\mu}(x) A_{c}^{\mu}(y): 
+ :u_{a}(x) \tilde{u}_{c}(y): - :\tilde{u}_{a}(x) u_{c}(y):]
\nonumber \\
+ 4h_{ab} h_{ab} D_{0}^{2}(x-y) {\bf 1} + c_{0} \delta (x-y) {\bf 1}
\nonumber \\
-2 f_{cab} h_{cd} [D_{0}(x-y) + c_{4} \delta (x-y)] 
[:A_{a\nu}(x) F_{b}^{\nu\mu}(x) A_{d\mu}(y): +
:A_{d\mu}(x) A_{a\nu}(y) F_{b}^{\nu\mu}(y): 
\nonumber \\
- :u_{a}(x) \partial_{\mu}\tilde{u}_{b}(x) A_{d}^{\mu}(y): 
- :A_{d}^{\mu}(x) u_{a}(y) \partial_{\mu}\tilde{u}_{b}(y): 
\nonumber \\
- :A_{a}^{\mu}(x) \partial_{\mu}\tilde{u}_{b}(y) u_{d}(y): 
- :u_{d}(x) A_{a}^{\mu}(y) \partial_{\mu}\tilde{u}_{b}(y):]
\nonumber\\
-2 f_{cab} h_{cd} {\partial \over \partial x_{\nu}} [D_{0}(x-y) + c_{5} 
\delta (x-y)] [ :A_{a\mu}(x) A_{b\nu}(x) A_{d}^{\mu}(y):
+ :A_{d}^{\mu}(x) A_{a\mu}(y) A_{b\nu}(y):
\nonumber \\
+ :A_{a\nu}(x) u_{b}(x) \tilde{u}_{d}(y):
+ :\tilde{u}_{d}(x) A_{a\nu}(y) u_{b}(y): ] \quad
\label{T2-generic}
\end{eqnarray}

Here
\be
D_{0,F}(x) \equiv D^{ret}_{0}(x) + D^{(+)}_{0}(x), \quad
\ee
is the usual Feynman propagator and
\be
D_{0,F}^{2}(x) \equiv D^{ret,2}_{0}(x) + D^{(+)}_{0}(x)^{2}, \quad
D_{0,F}^{3}(x) \equiv D^{ret,3}_{0}(x) + D^{(+)}_{0}(x)^{3}
\ee
are some generalisation of it.
\end{prop}

{\bf Proof:} The proof is elementary. The only things to be noticed is that if
one includes all the finite renormalisations compatible with power counting and
Lorentz covariance, one will get beside the contributions already appearing
into the formula above, some additional terms which however, can be proven to
be null in the adiabatic limit. More precisely:

$\bullet$
in the second term one could also include the expression:
$
\delta (x-y) :F_{a}^{\nu\mu}(x) F_{d\nu\mu}(y):
$
but this will produce, in the adiabatic limit, terms of the type
$
T^{(10)}
$
and
$
T^{(12)}
$
from (\ref{list}) which can be discarded on the same grounds as there;

$\bullet$
in the fifth term one could include 
$
[ c_{6}g_{\mu\nu} \square \delta (x-y) + 
c_{7} \partial_{\mu}\partial_{\nu} \delta (x-y)] 
:A_{a}^{\mu}(x) A_{d}^{\nu}(y):
$
but, in the adiabatic limit, they will lead to expressions of the type 
$
T^{(10-12)}
$
which can be discarded;

$\bullet$
in the sixth term one could include the finite renormalisation
$
{\partial \over \partial x_{\rho}} \delta (x-y) [\cdots ]
$
but in the adiabatic limit we get contributions of the type
$
T^{(10-13)}
$
which can be discarded;

$\bullet$
in the eighth term one should include
$
c_{8} \square \delta (x-y)
$
which gives, in the adiabatic limit, a contribution of the type
$
T^{(10)}
$
which can be ignored;

$\bullet$
finally, one could add at will local terms of the type
$
P(\square) \delta (x-y) \times {\bf 1}
$
with $P$ a polynomial of degree $> 1$, but these terms are null in the
adiabatic limit.
$\qed$

Of course, we do not have the guarantee that the expression (\ref{T2-generic})
leads to a well-defined operator on the factor space
${\cal H}^{r}_{YM}$;
in fact, one can show that this can happen if and only if some severe
restrictions are placed on the constants appearing in the expression of the
interaction Lagrangian \cite{AS}. For the sake of completeness we give below
this result.

\begin{thm}
The expression 
$T_{2}$
appearing in the preceding proposition leads, in the adiabatic limit, to an
well defined operator on
${\cal H}^{r}_{YM}$
if and only if: (a) the constants 
$f_{abc}$
verify the Jacobi identities:
\be
f_{abc}f_{dec} + f_{bdc} f_{aec} + f_{dac} f_{bec} = 0;
\label{Jacobi}
\ee
(b) the second order terms are not present i.e.
\be
h_{ab} = 0;
\ee
(c) the constants
$c_{i}, \quad i = 1,\dots,4$
are given by
\be
c_{1} = {i \over 2}, \quad c_{2} = 0.
\ee
\label{T2}
\end{thm}

{\bf Proof:}
(i) The ``brute force" computation can be avoided using the following trick 
\cite{DHKS1}. We have from the definition (\ref{d2}) of the distribution 
$D_{2}$,
the Leibnitz formula (\ref{Leibnitz}) and (\ref{k1}):
\begin{eqnarray}
d_{Q} D_{2}(x,y) = [d_{Q} T_{1}(x), T_{1}(y)] + [T_{1}(x), d_{Q} T_{1}(y)] =
\nonumber \\
i {\partial \over \partial x^{\mu}} [K^{\mu}_{1}(x), T_{1}(y)] -
i {\partial \over \partial y^{\mu}} [K^{\mu}_{1}(y), T_{1}(x)];
\label{split-d2-a}
\end{eqnarray}
this formula will be used to compute the left hand side. It is elementary to
see that the distribution 
$d_{Q} D_{2}(x,y)$
still has a causal support so it can be split causally.

On the other hand, the causal splitting (\ref{split}) implies 
\be
d_{Q} D_{2}(x,y) = d_{Q} R_{2}(x,y) - d_{Q} A_{2}(x,y).
\label{split-d2-b}
\ee

So, if we split causally the right hand side of the formula (\ref{split-d2-a})
without ruining Lorentz covariance and power counting, we get 
{\it a posteriori} valid expressions for the distributions
$d_{Q} R_{2}(x,y)$
and
$d_{Q} A_{2}(x,y)$.
Of course, in this way we do not get the most general expression for these
distribution because we have the possibility of finite normalisations. But the
arbitrariness for
$d_{Q} R_{2}(x,y)$
is {\it exactly the same} as the arbitrariness for
$d_{Q} T_{2}(x,y)$
which can be read from (\ref{T2-generic}). So, in this way, we get in a 
roundabout way, the most general expression for the distributions
$d_{Q} R_{2}(x,y)$
and
$d_{Q} A_{2}(x,y)$.

Finally, we see that (\ref{gi-Tn}) for $n = 2$ is equivalent to:
\be
lim_{\epsilon \searrow 0} \left. d_{Q} R_{2}(x,y)\right|_{Ker(Q)} = 0;
\label{gi-t2}
\ee
(one must use, of course the fact that we already have (\ref{gi-Tn}) for 
$n = 1$).
Imposing this condition on the expression determined in the way outlined above
will lead to the conditions from the statement.

(ii) A long, but straightforward computation gives us the following expression
for the first commutator appearing in (\ref{split-d2-a}):
\begin{eqnarray}
[K^{\mu}_{1}(x), T_{1}(y)] = D_{0}(x-y) T^{\mu}(x,y) + 
{\partial \over \partial x^{\nu}} D_{0}(x-y) T^{\mu\nu}(x,y) 
\nonumber \\
+ {\partial \over \partial x^{\nu}} D_{0,2}(x-y) \tilde{T}^{\mu\nu}(x,y) 
+ {\partial \over \partial x^{\nu}\partial x^{\rho}} D_{0}(x-y) 
T^{\mu\nu\rho}(x,y) 
+ D_{0}^{\mu\nu}(x-y) \tilde{T}_{\nu}(x,y) 
\nonumber \\
+ {\partial \over \partial x_{\mu}} D_{0}(x-y) T(x,y) 
+ {\partial \over \partial x_{\mu}\partial x_{\rho}} D_{0}(x-y) 
\tilde{T}_{\rho}(x,y)
\label{[k,t]}
\end{eqnarray}
where we have:
\begin{eqnarray}
T^{\mu}(x,y) \equiv
 f_{cab} f_{cde} 
[:u_{a}(x) F_{b}^{\nu\mu}(x) A_{d}^{\rho}(y) F_{e\rho\nu}(y): 
+ :u_{a}(x) F_{b}^{\nu\mu}(x) u_{d}(y) \partial_{\nu}\tilde{u}_{e}(y): ]
\nonumber \\
+ 2 f_{cab} h_{cd} [:u_{a}(x) F_{b}^{\nu\mu}(x) A_{d\nu}(y): 
- :A_{a\nu}(x) F_{b}^{\nu\mu}(x) u_{d}(y): 
+ :u_{a}(x) \partial^{\mu}\tilde{u}_{b}(x) u_{d}(y):
\nonumber \\
- :u_{d}(x) A_{a\nu}(y) F^{\nu\mu}_{b}(y):] 
+ f_{cab} h_{cd} :u_{d}(x) u_{a}(y) \partial^{\mu}\tilde{u}_{b}(y):
\nonumber \\
- 2 (h^{2})_{ab} [2 :u_{a}(x) A_{b}^{\mu}(y): 
+ :A_{a}^{\mu}(x) u_{b}(y):]
\end{eqnarray}
\begin{eqnarray}
T^{\mu\nu}(x,y) \equiv
- f_{cab} f_{cde} 
[:u_{a}(x) A_{b}^{\nu}(x) A_{d\rho}(y) F_{e}^{\rho\mu}(y): 
- :A_{a\rho}(x) F_{b}^{\rho\mu}(x) A_{d}^{\nu}(y)u_{e}(y): 
\nonumber \\
- :u_{a}(x) A_{b}^{\nu}(x) u_{d}(y) \partial^{\mu}\tilde{u}_{e}(y): 
- :u_{a}(x) \partial^{\mu}u_{b}(x) A_{d}^{\nu}(y) u_{e}(y): ]
\nonumber \\
- f_{cab} h_{cd} [2 :u_{a}(x) A_{b\nu}(x) A_{d}^{\mu}(y):
- :u_{d}(x) A_{a}^{\mu}(y) A_{b}^{\nu}(y): 
- :A_{d}^{\mu}(x) A_{a}^{\nu}(y) u_{b}(y):]
\end{eqnarray}
\be
\tilde{T}^{\mu\nu}(x,y) \equiv
{1\over 2} f_{cab} f_{cde} 
[:u_{a}(x) F_{b}^{\mu\nu}(y): + :F_{d}^{\mu\nu}(x) u_{a}(y): ]
\ee
\be
T^{\mu\nu\rho}(x,y) \equiv
f_{cab} f_{cde} 
:u_{a}(x) A_{b}^{\nu}(x) A_{d}^{\mu}(y) A_{e}^{\rho}(y): 
\ee
\be
\tilde{T}_{\nu}(x,y) \equiv f_{cab} f_{cde} :u_{a}(x) A_{d\nu}(y):
\ee
\begin{eqnarray}
T(x,y) \equiv
f_{cab} f_{cde} 
[:u_{a}(x) A_{b}^{\nu}(x) A_{d}^{\rho}(y) F_{e\rho\nu}(y): 
- :u_{a}(x) A_{b}^{\nu}(x) u_{d}(y) \partial_{\nu}\tilde{u}_{e}(y): 
\nonumber \\
+ {1\over 2} 
:u_{a}(x) u_{b}(x) A_{d}^{\rho}(y) \partial_{\rho}\tilde{u}_{e}(y): ]
+ f_{cab} h_{cd} [2 :u_{a}(x) A_{b\nu}(x) A_{d}^{\nu}(y):
+ :u_{a}(x) u_{b}(x) \tilde{u}_{d}(y):] 
\end{eqnarray}
and
\be
\tilde{T}_{\rho}(x,y) \equiv
- f_{cab} f_{cde} 
:u_{a}(x) A_{b\nu}(x) A_{d}^{\nu}(y) A_{e\rho}(y): .
\ee

(iii) We must split causally the distribution
$
{\partial \over \partial x^{\mu}} [K^{\mu}_{1}(x), T_{1}(y)].
$
The first contribution appearing in the formula (\ref{[k,t]}) leads to
$$
{\partial \over \partial x^{\mu}} [D_{0}(x-y) :T^{\mu}(x,y):] =
{\partial \over \partial x^{\mu}} D_{0}(x-y) :T^{\mu}(x,y): +
D_{0}(x-y) :{\partial \over \partial x^{\mu}} T^{\mu}(x,y):
$$
with an admissible splitting given by:
$$
{\partial \over \partial x^{\mu}} D^{ret}_{0}(x-y) :T^{\mu}(x,y): +
D^{ret}_{0}(x-y) :{\partial \over \partial x^{\mu}} T^{\mu}(x,y): =
{\partial \over \partial x^{\mu}} [D^{ret}_{0}(x-y) :T^{\mu}(x,y):]
$$
so the splitting can be done without affecting the total divergence structure.
The same analysis works for the next three contributions appearing into the
right hand side of (\ref{[k,t]}). A major difference appears in the last two
contributions. We have, for instance for the distribution:
\begin{eqnarray}
{\partial \over \partial x^{\mu}} 
[{\partial \over \partial x_{\mu}} D_{0}(x-y) :T(x,y):] =
\square D_{0}(x-y) :T(x,y): +
{\partial \over \partial x_{\mu}} D_{0}(x-y) 
:{\partial \over \partial x^{\mu}} :T(x,y): =
\nonumber \\
{\partial \over \partial x_{\mu}} D_{0}(x-y) 
:{\partial \over \partial x^{\mu}} :T(x,y): \nonumber
\end{eqnarray}
that a possible splitting is
\begin{eqnarray}
{\partial \over \partial x_{\mu}} D^{ret}_{0}(x-y) 
:{\partial \over \partial x^{\mu}} T^{\mu}(x,y): =
{\partial \over \partial x_{\mu}} \left[ D^{ret}_{0}(x-y) 
:{\partial \over \partial x^{\mu}} :T^{\mu}(x,y):\right] -
\nonumber \\
\square D^{ret}_{0}(x-y) :T^{\mu}(x,y): =
{\partial \over \partial x_{\mu}} \left[ D^{ret}_{0}(x-y) 
:{\partial \over \partial x^{\mu}} :T^{\mu}(x,y):\right] -
i \delta (x-y) :T^{\mu}(x,y): \nonumber
\end{eqnarray}
so in this case, beside the total divergence, a 
$\delta (x-y)$
contribution appears. The last term from (\ref{[k,t]}) is dealt with in the 
same way, only now we will get a term proportional to
${\partial \over \partial x^{\rho}} \delta (x-y)$
which one integrates by parts. Finally, we obtain (after some combinatorial
rearrangements) that a possible causal splitting of 
$
{\partial \over \partial x^{\mu}} [K^{\mu}_{1}(x), T_{1}(y)]
$
is:
\begin{eqnarray}
{\partial \over \partial x^{\mu}} [K^{\mu}_{1}(x), T_{1}(y)] =
{\partial \over \partial x_{\mu}} L_{\mu}(x,y) + 
\nonumber \\
i \delta (x-y) 
[(f_{cae} f_{cdb} - {1\over 2} f_{cab} f_{cde})
:u_{a}(x) F_{b\rho\nu}(x) A_{d}^{\nu}(x) A_{e}^{\rho}(x): 
\nonumber \\
- {1\over 2} f_{cab} f_{cde}
:\partial_{\rho}u_{a}(x) A_{b\nu}(x) A_{d}^{\nu}(x) A_{e}^{\rho}(x): 
\nonumber \\
+ (f_{cad} f_{cbe} - {1\over 2} f_{cab} f_{cde})
:u_{a}(x) u_{b}(x) A_{d}^{\rho}(x) \partial_{\rho}\tilde{u}_{e}(x): 
\nonumber \\
- 2 f_{cab} h_{cd} :u_{a}(x) A_{b\nu}(x) A_{d}^{\nu}(x):
- f_{cab} h_{cd} :u_{a}(x) u_{b}(x) \tilde{u}_{d}(x):] 
\label{split-comm}
\end{eqnarray}
where the explicit expression of the distribution
$L_{\mu}(x,y)$
is not important.

(iv) The decomposition of the distribution
$
{\partial \over \partial y^{\mu}} [K^{\mu}_{1}(y), T_{1}(x)]
$
can be obtained from the previous expression if we notice that in the causal
splitting 
$
D(y-x) = - D^{adv}(y-x) + D^{ret}(y-x)
$
the first term plays the r\^ole of the retarded part. So, one should make he
substitution 
$
D^{ret}(x-y) \rightarrow - D^{adv}(y-x) 
$
in the previous expression. Finally, one gets a possible causal splitting of
the distribution 
$d_{Q} R_{2}(x,y)$
from (\ref{split-d2-a}):
\begin{eqnarray}
d_{Q} R_{2}(x,y) =
i {\partial \over \partial x_{\mu}} R_{\mu}(x,y) 
- \delta (x-y) 
[(2 f_{cae} f_{cdb} - f_{cab} f_{cde})
:u_{a}(x) F_{b\rho\nu}(x) A_{d}^{\nu}(x) A_{e}^{\rho}(x): 
\nonumber \\
- f_{cab} f_{cde}
:\partial_{\rho}u_{a}(x) A_{b\nu}(x) A_{d}^{\nu}(x) A_{e}^{\rho}(x): 
+ (2 f_{cad} f_{cbe} - f_{cab} f_{cde})
:u_{a}(x) u_{b}(x) A_{d}^{\rho}(x) \partial_{\rho}\tilde{u}_{e}(x): 
\nonumber \\
- 4 f_{cab} h_{cd} :u_{a}(x) A_{b\nu}(x) A_{d}^{\nu}(x):
- 2 f_{cab} h_{cd} :u_{a}(x) u_{b}(x) \tilde{u}_{d}(x):] \nonumber
\label{Ret}
\end{eqnarray}

The arbitrariness of the decomposition can be read from (\ref{T2-generic}), so
the previous formula gets corrected to
\begin{eqnarray}
d_{Q} R_{2}(x,y) =
i {\partial \over \partial x_{\mu}} R_{\mu}(x,y) 
- \delta (x-y) 
[(2 f_{cae} f_{cdb} - f_{cab} f_{cde})
:u_{a}(x) F_{b\rho\nu}(x) A_{d}^{\nu}(x) A_{e}^{\rho}(x): 
\nonumber \\
- f_{cab} f_{cde} (1 - 2i c_{1})
:\partial_{\rho}u_{a}(x) A_{b\nu}(x) A_{d}^{\nu}(x) A_{e}^{\rho}(x): 
\nonumber \\
+ (2 f_{cad} f_{cbe} - f_{cab} f_{cde})
:u_{a}(x) u_{b}(x) A_{d}^{\rho}(x) \partial_{\rho}\tilde{u}_{e}(x): 
\nonumber \\
- 4 f_{cab} h_{cd} :u_{a}(x) A_{b\nu}(x) A_{d}^{\nu}(x):
- 2 f_{cab} h_{cd} :u_{a}(x) u_{b}(x) \tilde{u}_{d}(x): 
\nonumber \\
+ c_{ad} :\partial_{\mu}u_{a}(x) A_{d}^{\mu}(x): + 
c_{4} P_{4}(x) + c_{5} P_{5}(x)] \quad
\end{eqnarray}
where 
$
c_{ad} \equiv c_{2} f_{cab} f_{cde} + c_{3} (h^{2})_{ad}
$
and the polynomial
$P_{4}(x)$ 
(resp. $P_{5}(x)$)
contain Wick monomials of the type
$:A \partial A \partial u:, \quad :A A \partial \partial u:$
(resp. 
$:u A \partial \partial A:, \quad :u \partial \tilde{u} \partial u:$)
which do not appear in the other contributions.

Now we impose the condition (\ref{gi-t2}). It is not very hard to see that one
obtains: 
\begin{eqnarray}
lim_{\epsilon \searrow 0} \int_{\R^{4}} dx g_{\epsilon}(x)^{2}
[(2 f_{cae} f_{cdb} - f_{cab} f_{cde})
:u_{a}(x) F_{b\rho\nu}(x) A_{d}^{\nu}(x) A_{e}^{\rho}(x): 
\nonumber \\
- f_{cab} f_{cde} (1 - 2i c_{1})
:\partial_{\rho}u_{a}(x) A_{b\nu}(x) A_{d}^{\nu}(x) A_{e}^{\rho}(x): 
\nonumber \\
+ (2 f_{cad} f_{cbe} - f_{cab} f_{cde})
:u_{a}(x) u_{b}(x) A_{d}^{\rho}(x) \partial_{\rho}\tilde{u}_{e}(x): 
\nonumber \\
- 4 f_{cab} h_{cd} :u_{a}(x) A_{b\nu}(x) A_{d}^{\nu}(x):
- 2 f_{cab} h_{cd} :u_{a}(x) u_{b}(x) \tilde{u}_{d}(x): 
\nonumber \\
+ c_{ad} :\partial_{\mu}u_{a}(x) A_{d}^{\mu}(x): + 
c_{4} P_{4}(x) + c_{5} P_{5}(x)]|_{Ker(Q)} = 0.
\nonumber
\label{gi-2}
\end{eqnarray}

From the third term we get
$$
2 f_{cad} f_{cbe} - f_{cab} f_{cde} = (a \leftrightarrow b)
$$
which is equivalent to the Jacobi identity (\ref{Jacobi}) from the statement.
But, in this case the first terms is also null. We get from the rest
$
c_{1} = {i \over 2}, \quad c_{ad} = 0
$
and
$
f_{cab} h_{cd} = 0
$.
Because the constants
$f_{abc}$
correspond to a semi-simple Lie algebra, the Killing-Cartan form is strictly
positively defined and we get from the last condition that in fact we have
$h_{ab} = 0$. In this case the constants
$c_{3}, c_{4}, c_{5}$
should not be included in the expression of 
$T_{2}$
and we also have
$c_{2} = 0$.
$\qed$

\begin{cor}
The chronological product
$T_{2}$
is given by the following expression:
\begin{eqnarray}
T_{2}(x,y) = {i\over 2} f_{cab} f_{cde} \delta (x-y)
:A_{a\nu}(x) A_{b\nu}(x) A_{d}^{\mu}(x) A_{e}^{\nu}(x): 
\nonumber \\
- f_{cab} f_{cde} D_{0,F}(x-y) [
:A_{a\nu}(x) F_{b}^{\nu\mu}(x) A_{d}^{\rho}(y) F_{e\rho\mu}(y): 
+ :u_{a}(x) \partial_{\mu}\tilde{u}_{b}(x): 
+ :u_{d}(y) \partial^{\mu}\tilde{u}_{e}(y): 
\nonumber \\
- :A_{a\nu}(x) F_{b}^{\nu\mu}(x) u_{d}(y) \partial_{\mu}\tilde{u}_{e}(y): 
- :u_{d}(x) \partial_{\mu}\tilde{u}_{e}(x) A_{a\nu}(y) F_{b}^{\nu\mu}(y): ]
\nonumber \\
+ {1\over 2} f_{abc} f_{dbc} D_{0,F}^{2}(x-y) 
:F_{a}^{\nu\mu}(x) F_{d\nu\mu}(y):
\nonumber \\
+ f_{cab} f_{cde} {\partial \over \partial x_{\rho}} D_{0,F}(x-y) [
:A_{a\nu}(x) F_{b}^{\nu\mu}(x) A_{d\mu}(y) A_{e\rho}(y): 
- :A_{d\mu}(x) A_{e\rho}(x) A_{a\nu}(y) F_{b}^{\nu\mu}(y): +
\nonumber \\
:A_{a\mu}(x) A_{b\rho}(x) u_{d}(y) \partial^{\mu}\tilde{u}_{e}(y): 
- :u_{d}(x) \partial^{\mu}\tilde{u}_{e}(x) A_{a\mu}(y) A_{b\rho}(y): 
\nonumber \\
+ :A_{a}^{\mu}(x) \partial_{\mu}\tilde{u}_{b}(x) A_{d\rho}(y) u_{e}(y): 
- :A_{d\rho}(x) u_{e}(x) A_{a}^{\mu}(y) \partial_{\mu}\tilde{u}_{b}(y): ]
\nonumber \\
- f_{cab} f_{cde} {\partial^{2} \over \partial x_{\nu} \partial x_{\rho}}
D_{0,F}(x-y) :A_{a\mu}(x) A_{b\nu}(x) A_{d}^{\mu}(y) A_{e\rho}(y): 
\nonumber \\
-2 f_{abc} f_{dbc} D_{0,F,\mu\nu}(x-y) :A_{a}^{\mu}(x) A_{d}^{\nu}(y):
+ {1\over 2} f_{abc} f_{dbc} {\partial \over \partial x_{\rho}} 
D_{0,F}^{2}(x-y) [:A_{a\mu}(x) F_{d}^{\mu\rho}(y): 
\nonumber \\
- :F_{d}^{\mu\rho}(x) A_{a\mu}(y): 
- :u_{a}(x) \partial_{\rho}\tilde{u}_{d}(y): 
- :\partial_{\rho}\tilde{u}_{d}(x) u_{a}(y):]
\nonumber \\
- {1\over 3} f_{abc} f_{abc} \square D_{0,F}^{3}(x-y) {\bf 1} 
+ c_{0} \delta (x-y) {\bf 1}. \qquad
\label{T-2}
\end{eqnarray}

Here the constants
$f_{abc}$
are completely antisymmetric and verify the Jacobi identities and
$c_{0}$
is arbitrary.
The condition of unitarity can be satisfied if and only if 
$c_{0}$
is purely imaginary.
\end{cor}

{\bf Proof:}
The first statement follows without any problems from the expression
(\ref{T2-generic}) if we take into account the preceding theorem. The second
statement amounts to:
$$
T_{2}^{\dagger}(x,y) + T_{2}(x,y) = T_{1}(x) T_{1}(y) + T_{1}(y) T_{1}(x)
$$
which easily follows from:
\be
D_{0,F}(x) + \overline{D_{0,F}(x)} = D^{(+)}_{0}(x) + D^{(-)}_{0}(-x)
\ee
and similar relations for
$
D_{0,F}^{2},~D_{0,F}^{3}
$
and
$
D_{0,F,\mu\nu}
$.
$\qed$

\newpage
\subsection{Yang-Mills Fields coupled to Matter}

In this subsection we study the possibility of coupling Yang-Mills fields to
``matter". As previously, we suppose that we are given the Hilbert space of
``matter" 
${\cal H}_{matter}$
which should also be a Fock space. Then the coupled system is described in the
tensor product Hilbert space
${\cal F}_{YM} \otimes {\cal H}_{matter}$.
One can describe this Fock space, as in the end of subsection \ref{quant}, by
considering 
$
\tilde{\cal H}^{gh,r}_{YM} \equiv 
{\cal H}^{gh,r}_{YM} \otimes {\cal H}_{matter}
$
with the corresponding supercharge operator and forming the quotient
$Ker(Q)/Im(Q)$.

First, we have to generalise theorem \ref{T1}:
\begin{thm}
Let us consider the operator
\be
T_{1}(g) = \int_{\R^{4}} dx~ g(x) T_{1}(x)
\ee
defined on 
$\tilde{\cal H}^{gh,r}_{YM}$
with
$T_{1}$
a Lorentz-invariant Wick polynomial in 
$A_{\mu},~u~,\tilde{u}$
and the matter fields, verifying also
$\omega(T_{1}) \leq 4$.
Then 
$T_{1}(g)$
can induce an well defined non-trivial $S$-matrix, in the adiabatic limit, if
and only if it has the following form:
\begin{eqnarray}
T_{1}(g) = \int_{\R^{4}} dx~ g(x) 
\{ f_{abc} \left[ :A_{a\mu}(x)A_{b\nu}(x) \partial^{\nu} A_{a}^{\mu}(x): -
:A_{a}^{\mu}(x) u_{b}(x) \partial_{\mu} \tilde{u}_{c}(x): \right] +
\nonumber \\
h_{ab} \left[ : A_{a\mu}(x)A_{b}^{\mu}(x): -
2 :\tilde{u}_{a}(x) u_{b}(x):\right] + :A_{a}^{\mu}(x) j_{a\mu}(x): +
T_{1,matter}(x) \}.
\label{YM-1-matter}
\end{eqnarray}
Here 
$f_{abc}$
($h_{ab}$)
are completely antisymmetric (symmetric) constants,
$j_{a\mu}$
is a conserved current build only from the matter fields with
$\omega(j_{a\mu}) = 1,2,3$
and
$T_{1,matter}$
contains only the matter fields.
\label{T1-matter}
\end{thm}

The proof follows the lines of theorem \ref{T1} only we have to add to the list
of possible building blocks of 
$T_{1}$,
beside the list
$T^{(1)}-T^{(15)}$,
and
$T_{1,matter}$,
combinations of the type
\begin{eqnarray}
T^{(16)} = :A_{a}^{\mu} A_{b}^{\nu} A_{c}^{\rho} j^{abc}_{\mu\nu\rho}: \quad
T^{(17)} = :A_{a}^{\mu} A_{b}^{\nu} j^{ab}_{\mu\nu}:\quad
T^{(18)} = :A_{a}^{\mu} \partial^{\rho}A_{b}^{\nu} j^{ab}_{\mu\nu\rho}:
\nonumber \\
T^{(19)} = :A_{a}^{\mu} j^{a}_{\mu}:\quad 
T^{(20)} = :u_{a} \tilde{u}_{b} j^{ab}:\nonumber
\end{eqnarray}
with
$
\omega(j^{abc}_{\mu\nu\rho}) = 1, \quad 
\omega(j^{ab}_{\mu\nu}) = 1, 2\quad 
\omega(j^{ab}_{\mu\nu\rho}) = 1, \quad 
\omega(j^{a}_{\mu}) = 1, 2, 3,\quad 
\omega(j^{ab}) = 1, 2.
$
Moreover, we can impose the symmetry properties:
$$
j^{ab}_{\mu\nu} = j^{ba}_{\nu\mu}, \quad
j^{ab}_{\mu\nu\rho} = j^{ba}_{\nu\mu\rho}.
$$

Proceeding as before we end up with the expression from the statement.
Moreover, we have as before:
\begin{prop}
The expression for 
$T_{1}$
verifies the unitarity requirement if and only if we have:
\be
j_{a}^{\mu}(x)^{\dagger} = j_{a}^{\mu}(x)
\ee
and verifies the causality condition if and only if:
\be
[j_{a}^{\mu}(x), j_{b}^{\nu}(x) ] = 0,\quad (x-y)^{2} < 0.
\ee
\end{prop}

Now we consider a special case of matter fields, namely a collection of {\it
free} Dirac fields. This situation it is supposed to described quantum
chromodynamics.  First we have
\begin{prop}
Suppose that 
${\cal H}_{matter}$
is the Fock space of an ensemble of Dirac fields
$\psi_{A}$
of masses
$m_{A} \geq 0, \quad A = 1,\dots,N$.
Then the generic form of the current, verifying power counting, Lorentz
covariance, unitarity and conservation is:
\be
j_{a}^{\mu}(x) = 
:\overline{\psi_{A}}(x) (t_{a})_{AB} \gamma^{\mu} \psi_{B}(x): +
:\overline{\psi_{A}}(x) (t'_{a})_{AB} \gamma^{\mu} \gamma_{5} \psi_{B}(x):
\label{current}
\ee
where the numerical matrices
$t_{a},\quad t'_{a}, \quad a = 1,\dots r$
must be hermitian. Moreover, the first set of matrices can be exhibited into a
block diagonal structure (eventually after a relabelling of the Dirac fields)
and the masses corresponding to the same block must be equal. The second term
is admissible if and only if all the masses are null. Finally, the expression
of this current satisfy the causality condition from the preceding proposition.
\end{prop}

{\bf Proof:}
The generic form for the current from the statement follows from power counting
and Lorentz covariance. The hermiticity of the matrices is a consequence of the
unitarity condition from the preceding proposition. Finally, imposing the
conservation of this current we end up with the conditions:
$$
(m_{A} - m_{B}) (t_{a})_{AB} = 0, \quad (m_{A} + m_{B}) (t'_{a})_{AB} = 0,
\quad \forall A, B = 1,\dots N, \quad \forall a = 1,\dots r
$$
which lead to the conclusions from the statement. The verification of the
causality property is by straightforward computation.
$\qed$

We will restrict ourselves in the following to the case of non-zero masses. It
is clear from the block structure of the matrices
$t_{a}$
that, in fact, we do not loose the generality of the analysis if we take only
one block corresponding to a single mass $m$. Let us define some distributions
with causal support which will be needed in the next proposition and which do
appear in spinorial QED \cite{Sc1}:
\be
S^{(\pm)}(x) \equiv (i\gamma\cdot \partial + m) D^{(\pm)}_{m}(x),\quad
S(x) \equiv S^{(+)}(x) + S^{(-)}(x) = (i\gamma\cdot \partial + m) D_{m}(x),
\ee
where 
$D_{m}(x)$
is the Pauli-Jordan function for arbitrary mass $m$ defined similarly to 
$D_{0}(x)$ in (\ref{D0}) but the integral is done over the hyperboloid of mass
$X^{+}_{m}$. The causal splitting
$$
D_{m}(x) = D^{ret}_{m}(x) - D^{adv}_{m}(x) 
$$
induces a similar splitting for the distribution
$S(x)$
and in this way Feynman propagator can be obtained. We also have the
distribution with causal support:
\be
\Sigma(x) \equiv \gamma_{\mu} 
\left[ D^{(+)}(x) S^{(+)}(x) + D^{(+)}(-x) S^{(-)}(x) \right] \gamma^{\mu}
\ee
\be
P^{(\pm)}_{\mu\nu}(x) \equiv \pm Tr \left[ \gamma_{\mu} S^{(-)}(\mp x)
\gamma_{\nu} S^{(+)}(\pm x)\right], \quad
P_{\mu\nu}(x) \equiv P^{(+)}_{\mu\nu}(x) + P^{(-)}_{\mu\nu}(x)
\ee
and
\be
V^{(\pm)}(x) \equiv \pm g^{\mu\nu} D^{(\pm)}_{0}(x) P^{(\pm)}_{\mu\nu}(x),
\quad V(x) \equiv V^{(+)}(x) + V^{(-)}(x).
\ee

All these distribution can be split causally and preserving Lorentz covariance
and we will denote the corresponding retarded, advanced and Feynman
distributions in an obvious way.

Then we have the following generalisation of the formula (\ref{T2-generic}):
\begin{prop}
Suppose that 
$\psi_{A}, \quad A = 1,\dots,N$
are Dirac fields of mass 
$m > 0$
such that the current is vectorial. We also suppose that there is no
contribution 
$T_{1,matter}$
in the first order chronological product. Then, the generic form of the second
order chronological product is:
\begin{eqnarray}
T_{2}(x,y) = T^{YM}_{2}(x,y) 
\nonumber \\
- f_{abc} D_{0,F}(x-y) [ 
:A_{a\nu}(x) F_{b}^{\nu\mu}(x) j_{c\mu}(y): - :u_{a}(x)
\partial^{\mu}\tilde{u}_{b} j_{c\mu}(y): + (x \leftrightarrow y)]
\nonumber \\
- f_{abc} {\partial \over \partial x^{\nu}} D_{0,F}(x-y) [ 
:A_{a}^{\mu}(x) A_{b}^{\nu}(x) j_{c\mu}(y): - (x \leftrightarrow y)]
\nonumber \\
-2 h_{ab} D_{0,F}(x-y) [ 
:A_{a\mu}(x) j_{b\mu}(y): + (x \leftrightarrow y)] 
\nonumber \\
- D_{0,F}(x-y) :\overline{\psi}_{A}(x) (t_{a})_{AB} \gamma^{\mu} \psi_{B}(x)
\overline{\psi}_{C}(y) (t_{a})_{CD} \gamma_{\mu} \psi_{D}(y):
\nonumber \\
+ :\overline{\psi}_{A}(x) (t^{2}_{a})_{AB} \Sigma_{F}(x-y) \psi_{B}(y):
+ :\overline{\psi}_{A}(y) (t^{2}_{a})_{AB} \Sigma_{F}(y-x) \psi_{B}(x):
\nonumber \\
+ \delta (x-y) [:\overline{\psi}_{A}(x) M_{AB} \psi_{B}(x): +
: \overline{\psi}_{A}(x) \gamma_{5} M'_{AB} \psi_{B}(x):]
+ Tr(t^{2}_{a}) V_{F}(x-y) {\bf 1}
\nonumber \\
+ :A_{a}^{\mu}(x) A_{b}^{\nu}(y): \{
:\overline{\psi}_{A}(x) (t_{a}t_{b})_{AB} \gamma_{\mu} S_{F}(x-y) \gamma_{\nu}
\psi_{B}(y):
\nonumber \\
+ :\overline{\psi}_{A}(y) (t_{b}t_{a})_{AB} \gamma_{\nu} S_{F}(y-x)
\gamma_{\mu} \psi_{B}(x):
\nonumber \\
+ Tr(t_{a} t_{b}) [P_{F,\mu\nu}(x-y) + f_{1} g_{\mu\nu} \delta (x-y) + 
f_{2} \partial_{\mu}\partial_{\nu} \delta (x-y)]{\bf 1} \}.
\label{T2-YM}
\end{eqnarray}
\end{prop}

{\bf Proof:}
The first step is, as before, to compute explicitly the commutator
$D_{2}$. If we denote the pure Yang-Mills contribution (\ref{D2}) by
$D^{YM}_{2}(x,y)$
then we have by elementary computations:
\begin{eqnarray}
D_{2}(x,y) = D^{YM}_{2}(x,y) 
\nonumber \\
- f_{abc} D_{0}(x-y) [ 
:A_{a\nu}(x) F_{b}^{\nu\mu}(x) j_{c\mu}(y): - :u_{a}(x)
\partial^{\mu}\tilde{u}_{b} j_{c\mu}(y): + (x \leftrightarrow y)]
\nonumber \\
- f_{abc} {\partial \over \partial x^{\nu}} D_{0}(x-y) [ 
:A_{a}^{\mu}(x) A_{b}^{\nu}(x) j_{c\mu}(y): - (x \leftrightarrow y)]
\nonumber \\
-2 h_{ab} D_{0}(x-y) [ :A_{a\mu}(x) j_{b\mu}(y): + (x \leftrightarrow y)] 
\nonumber \\
- D_{0}(x-y) :\overline{\psi}_{A}(x) (t_{a})_{AB} \gamma^{\mu} \psi_{B}(x)
\overline{\psi}_{C}(y) (t_{a})_{CD} \gamma_{\mu} \psi_{D}(y):
\nonumber \\
+ : \overline{\psi}_{A}(x) (t^{2}_{a})_{AB} \Sigma(x-y) \psi_{B}(y):
- : \overline{\psi}_{A}(y) (t^{2}_{a})_{AB} \Sigma(y-x) \psi_{B}(x):
+ Tr(t^{2}_{a}) V(x-y) {\bf 1}
\nonumber \\
+ :A_{a}^{\mu}(x) A_{b}^{\nu}(y): \{
:\overline{\psi}_{A}(x) (t_{a}t_{b})_{AB} \gamma_{\mu} S(x-y) \gamma_{\nu}
\psi_{B}(y):
\nonumber \\
- :\overline{\psi}_{A}(y) (t_{b}t_{a})_{AB} \gamma_{\nu} S(y-x) \gamma_{\mu}
\psi_{B}(x):
+ Tr(t_{a} t_{b}) P_{\mu\nu}(x-y) {\bf 1} \}. \nonumber
\end{eqnarray}

Now one must proceed to the causal splitting of this distribution. The finite
renormalisations can affect the sixth, the seventh and the last term of the
expression from above. Lorentz covariance and power counting must be used now
to fix the generic form of the finite renormalisation. For the last term, the
structure of these contributions is clear. For the sixth term, we can add to
$\Sigma_{F}$ the finite renormalisation
$
\delta (x-y) \{:\overline{\psi}_{A}(x) 
[M_{AB} {\bf 1} \gamma\cdot\partial + \gamma_{5} \gamma\cdot\partial]
\psi_{B}(x): \}
$
but if we use Dirac equation for the free Dirac fields 
$\psi_{B}(x)$
we end up with the expression from the statement.
$\qed$

Finally, we check if the expression just derived induces an well-defined
operator on the physical space. 
\begin{thm}
In the conditions of the preceding proposition, the second order chronological
product 
$T_{2}$
induces, in the adiabatic limit, an well-defined operator on the physical space
${\cal H}_{total}$
if and only if, beside the conditions from theorem \ref{T2} we also have: 
\be
[t_{a}, t_{b}] = i f_{abc} t_{c}, \quad \forall a,b = 1,\cdots,r
\ee
and
\be
f_{1} = f_{2} = 0.
\ee
\end{thm}

{\bf Proof:}
As in the proof of theorem \ref{T2}, we compute the commutator
$[K^{\mu}_{1}(x), T_{1}(y)]$;
for simplicity, we denote the pure Yang-Mills contribution given by the formula
(\ref{[k,t]}) in a suggestive way by
$[K^{\mu}_{1}(x), T_{1}(y)]^{YM}$.
Then we get by direct computation:
\begin{eqnarray}
[K^{\mu}_{1}(x), T_{1}(y)] = [K^{\mu}_{1}(x), T_{1}(y)]^{YM} 
\nonumber \\
- f_{abc} {\partial \over \partial x^{\nu}} D_{0}(x-y) [ 
:j_{a}^{\mu}(x) A_{b}^{\nu}(x) u_{c}(y): 
+ :A_{b}^{\nu}(x) u_{c}(x) j_{a}^{\mu}(y):]
\nonumber \\
-2 h_{ab} D_{0}(x-y) [ :j_{a}^{\mu}(x) u_{b}(y): - :u_{b}(x) j_{a}^{\mu}(y):] 
+ f_{abc} D_{0}(x-y) :u_{a}(x) F_{b}^{\nu\mu}(x) j_{c\nu}(y): +
\nonumber \\
f_{abc} {\partial \over \partial x^{\mu}} D_{0}(x-y) 
:u_{a}(x) A_{b}^{\nu}(x) j_{c\nu}(y): 
- :u_{a}(x) \overline{\psi}_{A}(y) (t_{b} t_{a})_{AB} \gamma_{\rho}
S(y-x) \gamma^{\mu} \psi_{B}(x) A_{b}^{\rho}(x):
\nonumber \\
+ : u_{a}(x) \overline{\psi}_{A}(x) (t_{a} t_{b})_{AB} \gamma^{\mu}
S(x-y) \gamma_{\rho} \psi_{B}(y) A_{b}^{\rho}(y):
+ Tr(t_{a} t_{b}) P^{\mu\rho}(x-y) :u_{a}(x) A_{b\rho}(y): \nonumber
\end{eqnarray}

So, beside the causal decompositions analysed previously in theorem \ref{T2},
we have to decompose distributions of the type: 
\begin{eqnarray}
{\partial \over \partial x^{\mu}} : T_{\alpha}(x) \gamma^{\mu} S(x-y) 
T_{\beta}(y): =
:{\partial T_{\alpha}\over \partial x^{\mu}}(x) \gamma^{\mu} S(x-y) 
T_{\beta}(y): 
+ : T_{\alpha}(x) \gamma\cdot\partial S(x-y) T_{\beta}(y):
\nonumber \\
:{\partial T_{\alpha}\over \partial x^{\mu}}(x) \gamma^{\mu} S(x-y) 
T_{\beta}(y): 
- im : T_{\alpha}(x) S(x-y) T_{\beta}(y): \nonumber 
\end{eqnarray}
where in the last line we have used Dirac equation for the Pauli-Jordan
distribution. A possible splitting is given by 
\begin{eqnarray}
:{\partial T_{\alpha}\over \partial x^{\mu}}(x) \gamma^{\mu} S^{ret}(x-y)
T_{\beta}(y): - im : T_{\alpha}(x) S^{ret}(x-y) T_{\beta}(y): =
\nonumber \\
{\partial \over \partial x^{\mu}}[ :T_{\alpha}(x) \gamma^{\mu} S^{ret}(x-y)
T_{\beta}(y):] 
+ i : T_{\alpha}(x) (i\gamma\cdot\partial - m) S^{ret}(x-y) T_{\beta}(y): 
\nonumber \\
{\partial \over \partial x^{\mu}}[ :T_{\alpha}(x) \gamma^{\mu} S^{ret}(x-y)
T_{\beta}(y):] 
+ \delta (x-y) : T_{\alpha}(x) T_{\beta}(y): 
\end{eqnarray}
and a similar analysis is valid for the term
$
{\partial \over \partial x^{\mu}} : T_{\alpha}(x) \gamma^{\mu} S(y-x) 
T_{\beta}(y): 
$
with the result that only the $\delta$-term has here a minus sign. 

As a consequence, we have the following causal splitting 
\begin{eqnarray}
{\partial \over \partial x^{\mu}} [K^{\mu}_{1}(x), T_{1}(y)] = 
{\partial \over \partial x_{\mu}} L_{\mu} + \delta (x-y)^{YM} 
- i \delta (x-y) f_{abc} :u_{a}(x) A_{b}^{\nu}(x) j_{c\nu}(x):  
\nonumber \\
\delta (x-y) [ :u_{a}(x) \overline{\psi_{A}}(x) (t_{b} t_{a})_{AB} 
\gamma_{\rho} \psi_{B}(x) A_{b}^{\rho}(x): +
:u_{a}(x) \overline{\psi_{A}}(x) (t_{a} t_{b})_{AB} 
\gamma_{\rho} \psi_{B}(x) A_{b}^{\rho}(x): ]
\end{eqnarray}
where by
$\delta (x-y)^{YM}$
we mean the $\delta$-terms appearing in formula (\ref{split-comm}). The
treatment of the contribution
$
{\partial \over \partial y^{\mu}} [K^{\mu}_{1}(y), T_{1}(x)] 
$
follows the lines described in the derivation of the theorem \ref{T2} and we
obtain in the end the following modification of the possible choice for the
retarded component of the commutator $D_{2}$ (see (\ref{Ret})):
\begin{eqnarray}
d_{Q} R_{2}(x,y) = d_{Q} R_{2}(x,y)^{YM} + 2i \delta (x-y) 
:u_{a}(x) \overline{\psi_{A}}(x) \left( [t_{a}, t_{b}] - i f_{abc} t_{c}
\right)_{AB} \gamma_{\rho} \psi_{B}(x) A_{b}^{\rho}: 
\end{eqnarray}
where
$d_{Q} R_{2}(x,y)^{YM}$
is the expression given by (\ref{Ret}). The arbitrariness of this decomposition
becomes augmented (with respect to the similar one from the pure Yang-Mills
case) by the contribution:
\be
2i Tr(t_{a} t_{b}) \delta (x-y) 
\left[ f_{1} :\partial_{\mu}u_{a} A_{b}^{\mu}: +  f_{2} \left( 
:\partial_{\nu}\partial^{\mu}u_{a} \partial_{\mu}A_{b}^{\nu}: +
:\partial^{\mu} u_{a} \partial_{\mu}\partial_{\nu}A_{b}^{\nu}: \right)
\right] + {\rm div}
\ee
where by {\it div} we mean a total divergence.

The assertion from the statement follows now immediately from the condition 
(\ref{gi-t2}).
$\qed$

The meaning of the theorem is that the Dirac ``matter" fields should form a
multiplet for the Lie algebra with structure constants 
$f_{abc}$
i.e. they should transform according to some (finite dimensional)
representation of this algebra.

\begin{cor}
In the preceding conditions the expression of the second order chronological
product is:
\begin{eqnarray}
T_{2}(x,y) = T^{YM}_{2}(x,y) 
\nonumber \\
- f_{abc} D_{0,F}(x-y) [ 
:A_{a\nu}(x) F_{b}^{\nu\mu}(x) j_{c\mu}(y): - :u_{a}(x)
\partial^{\mu}\tilde{u}_{b} j_{c\mu}(y): + (x \leftrightarrow y)]
\nonumber \\
- f_{abc} {\partial \over \partial x^{\nu}} D_{0,F}(x-y) [ 
:A_{a}^{\mu}(x) A_{b}^{\nu}(x) j_{c\mu}(y): - (x \leftrightarrow y)]
\nonumber \\
- D_{0,F}(x-y) :\overline{\psi}_{A}(x) (t_{a})_{AB} \gamma^{\mu} \psi_{B}(x)
\overline{\psi}_{C}(y) (t_{a})_{CD} \gamma_{\mu} \psi_{D}(y):
\nonumber \\
+ :\overline{\psi}_{A}(x) (t^{2}_{a})_{AB} \Sigma_{F}(x-y) \psi_{B}(y):
+ :\overline{\psi}_{A}(y) (t^{2}_{a})_{AB} \Sigma_{F}(y-x) \psi_{B}(x):
\nonumber \\
+ \delta (x-y) [:\overline{\psi}_{A}(x) M_{AB} \psi_{B}(x): +
: \overline{\psi}_{A}(x) \gamma_{5} M'_{AB} \psi_{B}(x):]
+ Tr(t^{2}_{a}) V_{F}(x-y) {\bf 1}
\nonumber \\
+ :A_{a}^{\mu}(x) A_{b}^{\nu}(y): [
:\overline{\psi}_{A}(x) (t_{a}t_{b})_{AB} \gamma_{\mu} S_{F}(x-y) \gamma_{\nu}
\psi_{B}(y):
\nonumber \\
+ :\overline{\psi}_{A}(y) (t_{b}t_{a})_{AB} \gamma_{\nu} S_{F}(y-x)
\gamma_{\mu} \psi_{B}(x):
\nonumber \\
+ Tr(t_{a} t_{b}) P_{F,\mu\nu}(x-y) ].
\label{T2-YM-matter}
\end{eqnarray}
\end{cor}

\begin{rem}
One should expect that the arbitrariness of the expression written above
included in the ``mass" terms
$M_{AB}, \quad M'_{AB}$
will drop out, like in QED, if one goes to the third order of the perturbation
theory. 
\end{rem}
\newpage
\section{Conclusions}

The expressions (\ref{T-2}) and (\ref{T2-YM-matter}) are remarkable in the
sense that combining the general principles of perturbation theory with the
factorisation requirement (i.e. ``gauge invariance") we obtain fewer free
parameters that we would expect from the similar analysis performed on a model
without gauge invariance, like for instance, a scalar field theory with a
$:\phi^{l}:$
interaction;
however, these expressions have a common ``disease", namely they do not have,
in fact, adiabatic limit. Indeed, the constant term from (\ref{T-2}) exhibits
the usual logarithmic divergence and cannot be compensated like in the usual
treatment of infra-red divergences.  So, we do not have the stability of the
vacuum and rigorously speaking, these theories do not exists! This is the usual
problem with zero-mass theories and one can deal with it in two ways. One can
argue that from the physical point of view, the quarks are never free and as a
consequence a perturbation theory which describes asymptotic free particles has
no reason to exists. A more profound answer would be that the usual treatments
of the infra-red divergences are not sufficiently general and one should
proceed to a modification of the free Fock space taking into account the long
range nature of the interaction  mediated by zero-mass particles. Thus one
should look for some generalisation, in the Fock space formalism, of the
Dollard formalism of modified ``free" evolution.

There some further developments which will also be considered in future
publications, namely the same analysis in $2+1$ dimensions. In this way
Chern-Simons Lagrangian could be derived from general principles of
perturbation theory. A generalisation of the analysis from section 3 to massive
spin one particles is also necessary to complete the study of the electro-weak
interaction \cite{DS}, \cite{ASD3}. Finally, the same arguments should be also
implemented for the case of spin $2$ particles into an attempt to construct a
perturbative theory of gravity.

\end{document}